\documentclass[12pt]{article}

\usepackage[english]{babel}
\usepackage[utf8]{inputenc}
\usepackage[T1]{fontenc}
\usepackage{tabularx}
\usepackage{graphicx}
\usepackage{float}
\usepackage{enumerate}
\usepackage{placeins}
\usepackage{fancybox}
\usepackage{verbatim}
\usepackage{longtable}
\usepackage{url}
\usepackage{subfigure}
\usepackage{multirow}
\usepackage{amsfonts}
\usepackage{xcolor}
\usepackage{soul}
\usepackage{microtype}

\usepackage{authblk}
\usepackage{float}

\title{Adopting Road-Weather Open Data in Route Recommendation Engine}

\author[1]{Henna Tammia} 
\author[1]{Benjamin Kämä} 
\author[1]{Ella Peltonen} 
\affil[1]{\textit{Empirical Software Engineering in Software, Systems, and Services (M3S), University of Oulu. Pentti Kaiteran katu 1, Oulu, FI-90014, Finland}}
\date{}

\begin{document}

\section*{Highlights}
\textbf{Adopting Road-Weather Open Data in Route Recommendation Engine}\\
Henna Tammia, Benjamin Kämä, Ella Peltonen

\begin{itemize}
    \item Personalized route recommendations go beyond the shortest route but driver’s changing preferences.

    \item Open data services provide a vast amount of unused data from nationwide road networks.

    \item We utilize more than 2,300 real-time attributes from 1,814 road weather stations.

    \item We combine attributes into human-understandable information, such as visibility on the road.

    \item We showcase how open data enables personalized routes for drivers with different preferences.
\end{itemize}

\newpage

\maketitle

\begin{abstract}
Digitraffic, Finland's open road data interface, provides access to nationwide road sensors with more than 2,300 real-time attributes from 1,814 stations. However, efficiently utilizing such a versatile data API for a practical application requires a deeper understanding of the data qualities, preprocessing phases, and machine learning tools. This paper discusses the challenges of large-scale road weather and traffic data. We go through the road-weather-related attributes from DigiTraffic as a practical example of processes required to work with such a dataset. In addition, we provide a methodology for efficient data utilization for the target application, a personalized road recommendation engine based on a simple routing application. We validate our solution based on real-world data, showing we can efficiently identify and recommend personalized routes for three different driver profiles. 
\end{abstract}



\paragraph*{Keywords:} Open Data, Traffic Management, Recommendation Engines

\section{Introduction}


At its core, route planning is about finding an optimal route from the node of origin to a destination. This widely used functionality has been adapted and built upon for varied contexts and situations over the years. In the simplest types of route planning, the focus is on finding the shortest path in a static network modeled as a graph \cite{delling2009engineering,han2024privacy}. A multitude of different techniques have been used to find the shortest path in a graph, perhaps the best known of which is Dijkstra’s algorithm. Despite the high availability of different methods, it is to be noted that with each approach, there are trade-offs. Different methods can be faster, more robust, and more space efficient, but rarely all at once \cite{Bast2016}. Despite its name, the shortest path does not necessarily mean the shortest geographical distance between nodes. Heuristics defining the most suitable route can include cost types such as available fuel or power, the distance between optimal refueling or charging stations, types of roads preferred by the driver, e.g., new or older drivers may prefer to avoid busy highways, services, entertainment available on the road, e.g., restaurant options, and traffic load \cite{sadeghi2011real}. Road weather is the most relevant element for our work. To find the most suitable road, we considered issues such as avoiding roads with black ice, heavy rain or snowfall in the area, wind on bare seaside roads, known construction work, or heavy or long transportation items. Additionally, using only a single variable, be it time, distance, or something else, to represent the costs is insufficient as it fails to portray the complex reality of a real-world road network. 

It is also to be noted that the optimal route for a specific driver does not always correspond to the shortest or fastest path in the network, i.e., the global optimum. Beyond the traditional metrics of time and distance, energy consumption, and safety, different human preferences are the metrics that should be used to determine the optimality of a route in a truly personalized manner, as every driver is different \cite{taha2018route}. In static route planning, the network does not change, so the costs associated with each path remain the same. On the other hand, real-life road networks are not static but are subject to constant changes in traffic load, weather, and roadworks \cite{delling2009engineering}. In a report commissioned by the United States Department of Transportation, different events affecting road capacity are identified \cite{Dunn2006}. These events are broadly divided into planned and unplanned events. Planned events include maintenance and special events, while unplanned events cover incidents that cannot be predicted. The unplanned events cover events such as accidents, natural disasters, different types of emergencies, and adverse weather conditions. As any of these events may lead to congestion, they also prompt a need for alternative routes. The importance of choosing the optimal route in real time goes beyond ensuring user satisfaction. Optimized route planning can make the difference between life and death when emergency vehicles are involved.

A traffic management system (TMS) refers to systems that aim for congestion control and reducing other hazardous events on the road by controlling traffic lights, providing accident warnings, and supporting route planning and optimization \cite{de2017traffic}. The data collected via induction loop counters, cameras, and cellphone tower handover can be used to determine current traffic characteristics, such as vehicle speeds and traffic volume at specific points of the road network. This information can then be used for dynamic route planning \cite{djahel2014communications}. Congestion, i.e., traffic volume exceeding the capacity of the road disrupts the flow of traffic \cite{isa2015implementation}, increases travel times, pollution, and frustration of drivers unable to reach their destinations on time as route planning tends to stress towards global optima \cite{li2021towards}. Most current implementations of dynamic route planning revolve around congestion avoidance, that is, avoiding roads that are already jammed or can be predicted soon become overly crowded \cite{liebig2017dynamic}. 

However, there is more up-to-date information about the current state of the road network beyond just congestion. For example, Digitraffic, Finland's open road data interface \footnote{\url{https://www.digitraffic.fi/en/road-traffic/}}, provides access to nationwide road-weather sensors that can provide information about road and road-weather conditions vital for drivers to decide their preferred route of the day. This is extremely crucial in countries such as Finland, where rapidly changing weather can make roads undrivable, especially for less experienced drivers: heavy rain and winds during the autumn, black ice during autumn and spring, and snowstorms during long winter months. It is important to note that road weather, even if it overlaps, is not the same as typical weather and is less likely to be recognized by regular forecasts \cite{kangas2015roadsurf,national2004weather}. For example, after a snowstorm, the roads are cleared in a set prioritized order, and knowing when a certain road is suitable to drive requires external sensor data. Similarly, black ice and sudden gusts of winds can make roads unbearable even if the overall weather conditions are pristine.

In this paper, we argue that finding \textbf{personalized routes} is an essential factor for user satisfaction, safer driving, and avoiding congestion in the long term, as data-driven knowledge of the current state of the road network is more accessible to distribute for the different route-planning services and autonomous driving systems. To achieve this, we aim to produce a method that actively utilizes external open data to create real-time personable recommendations based on how the driver perceives the road conditions and how they affect his/her route decisions. In our first iteration, we plan to implement our system using the national road weather, traffic, and event information as a data source. To summarize, this paper provides the following contributions:

\begin{itemize}
    \item We present a comprehensive analysis of data collected from the national roadside weather measurement stations network. We discuss how it has to be preprocessed, fused, and cleaned to serve a purpose for a simple service, such as finding a suitable route under certain weather conditions.
    \item As an example use case, we showcase road weather, traffic, and event-aware recommendation systems for creating personalized preferred routes.
    \item We evaluate the recommendations with real Finnish Digitraffic open API data using three different weather situations for three different driver profiles. 
\end{itemize}





\section{Related Work}

To realize more optimal route planning in an ever-changing environment, various types of real-world data need to be acquired and used. Many modern road networks are already equipped with different monitoring technologies. This equipment includes cameras, traffic measurement systems, inductive loop detectors, and road weather stations. Accident detection and maintenance tracking are equally important factors in ensuring safe travels for road users, as are variable message signs and other ways of alerting the public of tumultuous conditions. Several applications have been proposed that utilize real-time road data. Even the same data can be used for vastly different purposes. In this section, we cover example use cases and previous literature on the following data features: roadside camera footage, road weather, weather forecast, traffic measurements, planned and unplanned events, and finally, maintenance tracking data.

\paragraph{Roadside Camera Footage} \label{relwo_cameras}
Recognizing congested roads is vital as it enables route planning around crowds \cite{shao2020estimation}. Vehicle counting, in general, is one of the most common uses of \textbf{CCTV cameras} on roads \cite{buch2011visionreview}. Roadside cameras aid in detecting incidents \cite{buch2011visionreview}, such as traffic accidents or violations \cite{anyang2016}, or congestion and traffic jams \cite{chakraborty2018traffic, kurniawan2018traffic}. Additionally, CCTV camera footage can be used for number plate recognition \cite{buch2011visionreview}, which can further be used to determine the volume and average speed of vehicles to assess the current traffic conditions \cite{zou2021cameratomap}. This is an especially valuable application for less trafficked roads \cite{ghosh2019accidentcamera}. 

\textbf{Roadside weather cameras} have a profound impact on roads. Their images can be used to determine the condition of the road and, therefore, whether maintenance, such as snowplowing and de-icing, needs to be dispatched \cite{jonsson2011classification, grabis2022maintenance, anyang2016}. Most cameras on roads are stationary, which introduces certain shortcomings. For example, they have limited visibility in certain weather conditions. To battle this, some works such as Chen et al. \cite{chen2022gocomfort} propose utilizing autonomous vehicles, which possess mobile cameras that can reach areas beyond the line of sight of stationary ones. They utilize vehicle-mounted cameras to capture images of potholes, which are then sent to a cloud for analysis and used to create a high-definition map of the potholes in the road network. Such a map can then be used to optimize route planning for passenger comfort.

\paragraph{Road-Weather Data Attributes}
Certain road-weather conditions are linked to significantly increased accident risks \cite{malin2019accident}. It is no surprise that historical weather data, together with accident records, have been used in creating road risk indices for assessing the safety of road segments \cite{li2015roadrisk}. Considering weather conditions is essential for route planning because heavy rain or snow can reduce road capacity up to 17\% or 27\%, respectively, in addition to reducing travel speeds. Rain can lower road capacity by 1\% - 17\% and speed by 1\% - 7\%, depending on the intensity of the rain. Similarly, snow reduces capacity by 3\%-27\% and speeds by 3\%-15\% \cite{agarwal2005weatherimpact}.

Real-time information about the current road condition is crucial for planning and dispatching maintenance operations. Appropriate use of maintenance operations in invaluable for safe driving, especially in winter. \textbf{Air and surface temperature, humidity, dew point, precipitation, and wind speed and direction} can be used in determining whether a road is dry, wet, covered in snow or ice, or if it is snowy with tracks on it \cite{jonsson2011classification}. Based on data from road weather stations and predefined formal rules, it is possible to automate decision-making regarding maintenance operations \cite{jokste2022latviamaintenance}. Furthermore, weather data has been used to predict the behavior of other travelers. For example, current weather data has been considered an explanatory factor for predicting traffic flow \cite{essien2021twitterweather} 
and travel speeds on roads \cite{kinoshita2017weather}. 
The specific type of weather data used differs vastly between applications. Some works have used only a few attributes, while others have used multiple. For example, Ye et al. \cite{ye2024temperature} only used temperature values from the OpenWeather API and route planning by Google Maps to enable travelers to select more comfortable routes. In contrast, Jonsson et al. \cite{jonsson2011classification} used air and surface temperature, humidity, dew point, precipitation, wind speed, and wind direction with camera images. 

\paragraph{Road-Weather Forecasts}
While tracking the current road conditions is invaluable, sometimes knowing the present state is not enough. If the route is only calculated based on the conditions prevalent at the beginning, the conditions may have changed significantly towards the end of a long trip. In such cases, using weather forecasts instead of current weather forecasts can provide more appropriate route guidance throughout their trip. Using the predicted precipitation, wind, and surface temperature to estimate travel times of road segments as was done by Litzinger et al. \cite{litzinger2012weatherroute} could help plan out even more optimal routes and also estimate overall travel times with higher accuracy. Current road weather can be used as a basis for road weather to enable more localized forecasting. Road weather stations can predict adverse weather conditions based on predetermined rules. This makes it possible to warn road users of dangerous driving conditions~\cite{tomas2016forecasting}. 

\paragraph{Traffic Measurements}
 Broadly speaking, traffic measurements include \textbf{vehicle speeds and traffic volume}. The data can be used for route planning, parking management, and short-term traffic predictions. Various types of sensors are used to gather the data \cite{djahel2014communications}. One type of such sensors is a loop detector, which can be used to measure traffic speed and volume to determine prevailing traffic status \cite{anyang2016}. Loop detectors make it possible to calculate the speed of passing vehicles as well as traffic volume. This means that they can be used for detecting traffic jams, accidents, and even vehicles driving in the wrong direction \cite{jayasheelan2016tms}. 
Similarly, SCATS sensors have been used to gain access to traffic flow and vehicle speeds, which can then be used to create traffic predictions with the help of machine learning methods. The predictions can then be used as costs in route planning \cite{liebig2017predict}. Sensors can also be mobile by deploying them on taxis \cite{anyang2016}. Should congestion be detected or predicted, road vehicles can be rerouted to avoid it \cite{de2016icarus}.

The importance of monitoring the current traffic situation is highlighted in cases where time is of the essence. Such situations can include route planning for emergency services that require the fastest possible route and cannot afford to get stuck in congestion. Tracking the average travel speeds of each road segment is one way of using traffic data to realize optimal route planning in such a case \cite{al2018accident}. Regular drivers can have varying preferences on whether they want the fastest route based on the current traffic situation or perhaps the one that is generally the fastest based on historical data \cite{jakimavieius2010vilnius}.

\paragraph{Planned events}
Different sorts of events can massively affect the flow of traffic. These incidents can either be classified as planned or unplanned events. Planned events can include things like concerts and other social gatherings or construction work \cite{Dunn2006}. Such incidents also include heavy and oversized transports, requiring a special permit \cite{elyAbnormalTransport}. Predicting the congestion caused by planned events can be a challenge. Extensive data mining has been proposed to predict traffic congestion brought on by planned events based on the characteristics of the event as well as traffic flow and zone data \cite{fernando2019plannedevents}. Many specially planned events like sports events and concerts are characterized by two distinct waves of congestion, one before the event and one after. In such cases, it is possible to use the historical data of similar events and traffic measurements of the first wave to predict the second wave of congestion \cite{kwoczek2014predictingPSE}. Avoiding the affected roads would ensure smoother travel as it would prevent the user from getting stuck in traffic.

\paragraph{Unplanned events}
Unplanned traffic events include incidents that cannot be predicted, such as road accidents, and emergencies brought on by adverse weather \cite{Dunn2006}.
Road accidents involving different types of vehicles can present very different needs for route planning. For example, accidents involving hazardous chemical transport can have a dramatic effect on roads beyond the one where the accident took place \cite{li2024chemicalemergency}.
Generally, there are two approaches to dealing with accidents. The first is pre-emptive and entails assessing the risk associated with each road segment based on historic data \cite{li2015roadrisk} and using the risk as a cost for route planning \cite{liao2022accident}.
The second approach focuses on handling accidents as they happen. Accident detection plays a vital role in this approach. CCTV cameras can be utilized in identifying incidents on roads, and other road users can also report them \cite{anyang2016}. Additional sensors on board vehicles themselves can also be utilized in detecting accidents, for example, by detecting the deployment of airbags \cite{sherif2014real}, or via OBD2 \cite{de2015garuda}. Vehicles can then alert a control center so emergency personnel can be dispatched \cite{sherif2014real}, or a system can alert other road users via VANET to avoid the affected areas and provide them with new routes \cite{de2016icarus}.
The Traffic Message Channel (TMC) is another way of warning driver of incidents that can potentially affect their travels. It aims to provide road users with up-to-date information about any relevant events on the road \cite{davies1989TMCstandards}. TMC can also relay alerts about oncoming adverse weather to users \cite{tomas2016forecasting}. Taking TMC alerts into account in route planning makes avoiding areas affected by different disruptions possible even before congestion begins.

\paragraph{Maintenance Tracking}
Road maintenance can reduce road capacity and cause congestion \cite{Dunn2006}. Despite this, beyond mentions of potential causes for congestion to be avoided, maintenance activities are often ignored in route planning. This is evident by the limited literature on the subject.
Automatic vehicle location systems can be used to track the location of the maintenance vehicle and any relevant info on the maintenance that was conducted (time, type, etc.). This information can be used as input for further maintenance decision-making and reporting \cite{grabis2022maintenance, kocianova2015maintenance}. Maintenance vehicles could also potentially be fitted with sensors to monitor road conditions. \cite{kocianova2015maintenance}. Maintenance vehicles often travel at slower speeds or cause other disruptions to the traffic flow, which can cause significant inconveniences. Therefore, it is advised to avoid road segments with active maintenance operations. 
\section{Digitraffic Road-Weather Data, Attributes, and Preprocessing} \label{roaddata}
\begin{table*}[ht]
    \centering
    \scriptsize
    \begin{tabular}{ | p{1.5cm} | p{1.5cm} | p{1.5cm} | p{7.5cm} |} 
        \hline
         \textbf{Topic} & \textbf{Shortest Update Interval} & \textbf{MQTT/ JSON} & \textbf{Content} \\
         \hline
        \textbf{Road-weather cameras} & 10 mins & No/Yes & .jpeg images from camera stations. It can detail different directions of the road, the landscape, or the road surface. \\
         \hline
         \textbf{Current road-weather} & real-time & Yes/Yes & Sensor values (differ between stations): wind speed, humidity, general conditions, friction, the amount of ice/snow/water/salt on the road, visibility, dew point, frost point \\
         \hline
         \textbf{Road-weather forecast} & 5 mins & No/Yes & Forecasts for the next 12 hours. Contains: road and air temperatures, wind speed and direction, road condition, daylight status, forecast reliability, and conditions according to which the forecasts are made (precipitation, road, wind, freezing rain, visibility, friction conditions, and winter slipperiness). \\
         \hline
         \textbf{TMS data} & real-time & Yes/Yes & Average speeds free flow percentages, and number of vehicles detected in the past 5 minutes and the past hour. \\
         \hline
         \textbf{Traffic messages} & real-time & Yes/Yes & Available as DATEX II or JSON. Provides location as well as details of accidents, road work, adjusted weight limits, and exempt transport: situation type, announcement type, version, release time, version time, contact info, announcement in natural languages containing title location, features, estimated duration, sender, and any additional information. \\
         \hline
         \textbf{Variable signs} & real-time & No/Yes & Type, sign location, the value displayed, a time when the value has become valid, whether the value was changed automatically or manually, reliability of the sign, and the specific text displayed on each row of the sign.\\
         \hline
         \textbf{Road maintenance} & real-time & Yes/Yes & Start and end times of maintenance tasks, task type, latest location. \\
         \hline
    \end{tabular}
    \caption{Topics of data available through the Digitraffic API. The update intervals differ between the MQTT and JSON APIs even if both are available, with 1 minute being the shortest possible interval for JSON. 
    }
    \label{tab:datatable}
\end{table*}

\color{black}
Digitraffic is an open data service provided by Fintraffic \footnote{\url{https://www.digitraffic.fi/en/road-traffic/}}, a traffic management entity owned by the Finnish Ministry of Transport and Communications. It provides traffic data in real-time through open APIs. The data includes images from road weather cameras, current weather measurements, weather forecasts, traffic measurements and alerts, data displayed on variable message signs, and maintenance tracking (see Table \ref{tab:datatable}). Additionally, measurement site metadata is updated twice daily. A similar service can be found in New South Wales, Australia, called Live Traffic NSW \footnote{\url{https://www.livetraffic.com/}}, which provides information about incidents and adverse weather, rest areas, car parks, as well as Google-produced traffic flow status and live traffic camera images. The service produces a list of "hazards" along the route. 


\textbf{The challenge:} An open data service like Digitraffic provides just the interfaces for fetching the data. Thus, heavy preprocessing steps are required to clean and format the data for any purpose. In this paper, we aim to use road weather data to enrich route planning. To do this, we first investigated the full extent of the data available through the road traffic API, after which we moved on to defining weight vectors for a road network graph. 

\textbf{Weather cameras:} Of the 761 camera locations, one has been temporarily removed, leaving us with 760 active sites. Each site can have anywhere between 1 and 7 "presets," i.e., directions where the camera is aimed. The most common number of presets per location is three: one in each direction of traffic and one that is aimed at the road surface itself. Pictures are updated every 10 minutes, and it is possible to retrieve all pictures taken within the past 24 hours. While camera footage could be used for various purposes, including road surface state monitoring, this information can be retrieved through the weather API. That being said, future iterations utilizing road weather camera images and data from weather stations should be considered, as there are cameras in many locations without weather stations. This would help retrieve accurate data from a wider geographical area, in particular areas where there are no weather stations. However, in this paper, we focus on road weather attributes available at road weather stations, which will be discussed next. 

\textbf{Road weather stations:} In July 2024, 529 weather stations gathered data actively, while four were temporarily removed. Interestingly enough, 58 of these \textit{road} weather stations are located at airport runways. The state on the surface of a runway cannot be directly linked to that on ordinary roads, even if they are located nearby. Excluding these stations leaves us with 471 active road weather stations. 
Most stations are labeled as one of two station types: ROSA or RWS\_200. These coincide with the names of two types of road weather stations distributed by Vaisala, ROSA, and RWS200, of which ROSA has been discontinued \cite{vaisalaUpgradeROSA}. As the stations are commercial products, information about them is limited, and documentation is, at times, scarce.  While some information from Vaisala is available about the sensors and the values they produce (such as in the station product catalog \cite{RWS200ProductCatalog}), this information is limited and, at times, hard to link back to the attributes present in the Digitraffic API. This makes understanding the API output a challenge. Attributes carrying categorical information, such as weather conditions and road surface states, are particularly problematic, as despite these classes being listed with their names in the specific sensor documentation, the values returned for the attributes are numerical instead of strings. In cases where no distinction about the sensor used is made in the metadata, in addition to no value descriptions, it is impossible to know which states the attribute tries to communicate.

The road weather stations employ various sensors to gather data about their surroundings. The data can be fetched either through a JSON API or with an MQTT WebSocket. The attributes are given identifiers between 1 and 226. However, metadata can only be fetched for 148 sensors (with \verb|/api/weather/v1/sensors|). It includes identifiers, names, units, accuracy, and sensor value descriptions. The level of detail in the documentation varies greatly between sensors. In some cases, the descriptions are very in-depth, for example, by providing explanations for enumerate values, while others are missing basic things like units. Additionally, some sensors seem to measure the same thing. For example, there are four near-identical attributes for road surface temperature: \verb|TIE_1, TIE_2, TIE_3, TIE_4| (id 3, 5, 101, 111).
Different sensors likely measure temperature on different lanes; however, this is not explicitly stated. It could also be that the multiple sensors guarantee reliability, which would be very important as the temperature of the road surface is a vital piece of information for many applications of road weather data. We categorize attributes thematically into four groups: road surface conditions, visibility, environmental factors (e.g., extreme temperatures and high winds), and those that do not influence driving conditions. By combining the previously discussed duplicate variables into one attribute, we were left with 86 attributes (see Table \begin{table}[]
    \centering
    \scriptsize
    \begin{tabular}{|p{4cm}|p{7cm}|c|} \hline
         \textbf{Data item} & \textbf{Value} & \textbf{Weight} \\ \hline 
         \multirow{9}{*}{Road surface condition} &  Dry & 0  \\
    & Moist &  0.125  \\
    & Wet & 0.25  \\
    & Most likely moist and salty & 0.375  \\
    & Wet and salted & 0.5  \\
    & Frost & 0.625  \\
    & Ice & 0.75  \\
    & Snow & 0.875 \\
    & Slush & 1 \\ \hline

    \multirow{2}{3.5cm}{Surface temperature difference to freezing point} & $\geq$ 5°C & 0.0 \\
    & $\leq$ 0°C & 1.0 \\ \hline
    
    \multirow{2}{4cm}{Surface temperature difference to dew or frost point} & $\geq$ 5°C & 0.0 \\
    & $\leq$ 0°C & 1.0 \\ \hline
    

    \multirow{2}{*}{Friction} & 0.82 $\mu$ & 0.0 \\
    & 0.09 $\mu$ & 1.0 \\ \hline

    \multirow{2}{*}{Moisture}  & 0.0 mm & 0.0  \\
    & $\geq$ 7.0 mm & 1.0 \\ \hline

    \multirow{2}{*}{Snow depth}  & 0.0 cm & 0.0  \\
    & $\geq$ 10.0 cm & 1.0 \\ \hline  

    \multirow{3}{*}{Road temperature} & $\leq$ -20 or $\geq$ 5 & 0.0 \\
    & -2 & 1.0 \\ \hline
    
    \multirow{2}{*}{Relative humidity} &  0 \% & 0.0 \\
    & 100 \% & 1.0 \\ \hline

    \multirow{2}{*}{Precipitation intensity} & 0 mm/h & 0.0 \\
    & $\geq$ 10 mm/h & 1.0 \\ \hline
    
    \multirow{10}{*}{Precipitation type} &  Dry weather & 0.0 \\
    &  Weak precipitation, type cannot be determined & 0.111 \\
        & Drizzle & 0.222 \\
        & Rain & 0.333 \\
        & Wet sleet & 0.444 \\
        & Sleet & 0.556 \\
        & Hail & 0.667 \\
        & Freezing drizzle & 0.778 \\
        & Snow & 0.889 \\
        & Freezing rain & 1.0 \\ \hline

    \multirow{2}{*}{Visible distance} & $\geq$ 10000 m & 0.0 \\
    & 0 m & 1.0 \\ \hline

    \multirow{2}{3.5cm}{Air temperature difference to dew or frost point} & $\geq$ 5°C & 0.0 \\
    & $\leq$ 0°C & 1.0 \\ \hline

    \multirow{2}{*}{Air temperature} & 14°C & 0.0 \\
    & $\leq$ -30°C or $\geq$ 27°C & 1.0 \\ \hline

    \multirow{2}{*}{Average wind} & 0 m/s & 0.0 \\
    & $\geq$ 21 m/s & 1.0 \\ \hline

    \multirow{2}{*}{Maximum wind} & 0 m/s & 0.0 \\
    & $\geq$ 31.5 m/s & 1.0 \\ \hline

    \multirow{2}{*}{Free flow speed} & $\geq$ 100\% & 0.0 \\
    & 0.0\% & 1.0 \\ \hline

    \multirow{2}{*}{Vehicles-to-capacity ratio} 
    & 0\% & 0.0 \\
    & $\geq$ 100\% & 1.0 \\ \hline

    \multirow{4}{*}{Road work} 
    & No road work & 0.00 \\
    & Low severity & 0.33 \\
    & High severity & 0.66 \\
    & Highest severity & 1.00 \\ \hline

    \multirow{2}{*}{Accident} 
    & No accidents & 0.0 \\
    & Accident present & 1.0 \\ \hline
    
    \end{tabular}
    \caption{An overview of all attributes used in the work. Note that for items that have weights only listed as 0.0 and 1.0, the weights of values between the reported ones are scaled.}
    \label{tab:weights}
\end{table}
). The following sections detail the eventual classification of the attributes and their roles in the weight vector definition necessary for the routing algorithm. 

\subsection{Road Surface Conditions}
The condition of the road surface is one of the most critical factors affecting driving, handling the car, and braking distances. Several attributes offer classifications of the conditions on the surface of the road. These attributes are perhaps the most expressive regarding information conveyed by a single attribute. Their usefulness is further enforced by the fact that they are among the most common attributes actively updated by weather stations. One attribute classifies the conditions as dry, moist, wet, most likely moist and salty, wet and salted (salting agent has been spread on the road), frost, ice, snow, and slush (see Table \ref{tab:weights}). Slush is the most dangerous surface state as it is linked to a significantly higher accident risk than most other states, including dry, moist, frosty, and even icy roads \cite{malin2019accident}. Therefore, slush was considered the worst one of the possible surface states, and given the weight of 1. The dry road surface was likened to the absence of environmental hindrances. Therefore, it received a weight of 0. Moist was rated at 0.125, wet at 0.25. Their salted counterparts were ranked slightly higher, with moist and salty at 0.375 and wet and salted at 0.5. The presence of salting agents implies a higher risk for the formation of ice and slush. Malin et al. \cite{malin2019accident} found a higher risk for accidents associated with snowy rather than icy roads, perhaps since snowfall can start suddenly. Some roads might be icy throughout winter and, therefore, represent a continuous state rather than abnormal conditions. A frosty road is associated with a higher relative risk of accidents than moist roads but lower than snowy or icy roads. Therefore, frosty roads were assigned the weight of 0.625, icy 0.75, and snowy 0.875.

Ranking the road surface conditions is difficult, even from the perspective of accident minimization, as different combinations of surface and precipitation conditions contribute to very different relative risks. For example, a slushy road with no precipitation can be related to a higher risk of accidents than an icy one. Still, with heavy precipitation, an icy road becomes more than twice as accident-prone as a slushy road with similar precipitation \cite{malin2019accident}. The specific combination of factors can be extremely dangerous even if, separately, these two factors have a much smaller effect on driving conditions. An example of such a case is light snowfall combined with temperatures below -7°C, as the two factors make for slippery roads that cannot be salted \cite{FMIconditions}. With this in mind, this ranking of road surface states is an approximation.

Two other attributes describe the road surface conditions. However, these attributes provide numerical values with no value descriptions, and as such, we are unable to decipher the condition communicated by these attributes. Therefore, the above is the only surface condition attribute we will consider for the weight vectors. The weather stations measure the depth of moisture in general, or more specifically, water, snow, and ice on the road, all provided via their own attributes in millimeters. There are at least three sensors that can provide this data or a portion of it: the DSC optical sensor, the DRS511 embedded sensor, and a water level and snow depth sensor by Campbell Scientific \cite{RWS200ProductCatalog}. This makes determining the value ranges challenging. The maximum values of the DSC sensor appear to be the smallest at 2.00 mm, and the DSR511 can detect water levels up to 7mm, while the Campbell Scientific SR50A does not give a definitive limit but can output values up to 9999 mm. 



Additionally, a thicker layer often means more challenging driving conditions. More water lessens friction, as does a thick layer of snow - enough fresh snow on the road surface might prevent the wheels from touching the solid road surface. Ice is a more complicated form of moisture, as it does lessen friction, and thick layers of ice can be linked to deep ruts that make steering challenging. However, a thin layer of slippery ice on the road surface can arguably be even more dangerous as it can be harder to detect. Regardless, as the attribute conveying the unspecified moisture level is a separate and far more common attribute than the more specific ones (except snow depth, which is equally common), we had to work under the assumption that a thicker layer denotes further deteriorated conditions. The maximum layer thickness proved equally challenging, as we only knew that this value was produced by a "surface sensor," but not which one – as the DRS511 had a lower maximum than the Campbell sensor, we elected to follow that one. This set the maximum weight of 1 at 7mm, with 0mm being weighted at 0 and 3.5mm at 0.5.
The only specific moisture attribute that was commonly available was the snow depth. This is presumably provided by a different station, as instead of millimeters, the value is given in centimeters. As the SR50A documentation,  \footnote{\url{https://s.campbellsci.com/documents/cr/manuals/sr50a.pdf}}, we could find did not contain explicit value ranges, except for when the value is reported in millimeters for which the maximum value is 9999 mm. However, even in Finland, nearly 10 meters of snow on the road surface is not likely. On the other hand, snowfall of 10 to 15 cm overnight is certainly not unheard of, and it can cause difficulties for drivers. We reasoned that once the snow depth reaches 10 cm, the relative impact of additional snow is considerably smaller than below 10 cm. Therefore, we chose 10 cm as our maximum value for the snow depth, with 0 cm being the logical minimum value. 

As was indicated, friction on the road surface is an important factor that greatly affects the driving experience. The API provides this data as a value between 0.09 – 0.82µ, 
which greatly complements our understanding of the current state of the road surface conditions provided by the condition categories. Less friction can be make driving unsafe, therefore should the friction value be 0.09, the weight of the friction factor will be 1. A friction of 0.82µ will lead to a weight of 0. The values in between will be scaled evenly, e.g., the friction value of 0.455µ will result in a weight of 0.5.

In winter, friction can be improved using salting agents on the road surface. The Digitraffic API provides data on the number of agents present on the road surface and the concentration of said agents. The amounts are given as grams per square meter, while the concentration is given as grams per liter.
However, excessive amounts of road salt do not guarantee the best possible driving conditions. Road salt is notorious for corroding cars; its effectiveness worsens when the temperature drops below -6 degrees Celsius, and in a worst-case scenario, excessive salting in sub-optimal conditions can lead to the formation of even more slippery ice, should the melted water freeze again \cite{vaylavirastoTalvikunnossapito}. It can also promote slush formation, which Malin et al. \cite{malin2019accident} found among the most detrimental factors to road safety. All that being said, using the amount and concentration of salt on the road to assess the driving conditions is tricky and often heavily dependent on the driver's preferences. The friction and surface conditions resulting from the use of salt are generalizable. Therefore, they were used instead of the salt attributes for the weight creation.

Malin et al. also found that when the temperature of the road surface drops from +2°C to -2°C, the risk of accidents increases rapidly, after which the risk slowly decreases but remains elevated compared to temperatures above freezing \cite{malin2019accident}. The API provides the road surface temperature at an accuracy of one decimal. A single station can have up to five road temperature readings, and some stations can even provide the temperature 30 cm below ground. 
While extremely high road surface temperature can soften the road surface and cause other issues, in the Finnish context, this extreme is negligibly rare and primarily omitted from our work. Instead, our focus was on the temperatures identified by Malin et al. \cite{malin2019accident} as the most dangerous ones. The relative accident risk peaks at -2°C; therefore, we will consider this our maximum value and assign the weight of 1. The minimum values for the high and low are not symmetrical to reflect the relative risks. On the positive end of the temperature scale, the risk only begins to elevate at +2°C. To include a small safety buffer, we set the minimum weight of 0 at the temperature of +5°C and higher. The other minimum was set at -20°C to account for the slow decline of the risk index with the falling temperatures with the values in between scaling proportionally \cite{malin2019accident}.

The Digitraffic API includes many other road temperature-related attributes that focus more on the accumulation of potential moisture on the road than the temperature itself. The weather stations monitor the dew point of the road, as well as the freezing and frost points. While these limit values are important for route planning, they are only meaningful compared to the current temperature. 
API also offers attributes that provide the difference between the limit value and the current temperature of the road and air for both dew and frost points. In the context of the road surface, we used the ones that compared the road surface temperature rather than air. While the temperature is above freezing, the values of the dew point and frost point are usually the same, but once the temperature drops below freezing, they start to differ. As the attributes conveying the dew point are commonly available when the temperature is above freezing, we used it exclusively for weight creation. Once below freezing, we used the frost point attributes, except for weather stations that did not carry this attribute, in which case we only used the dew point attributes. The dew point is used when the temperature is above freezing, and the frost point is used when it is below, based on a recommendation from Vaisala's documentation \footnote{\url{https://www.vaisala.com/sites/default/files/documents/VIM-GLO-EN-HUM-Dew_point_frost_point_selection_\%20B212716EN-A.pdf}}.


Notable conditions only emerge once the difference between the dew or frost point and the current temperature becomes very small. For the most part, if there is a difference, the value should not affect route planning. If there is no difference or negative difference, the weight is set to 1. To account for measuring errors and to add a small safety buffer, as conditions might change relatively rapidly, the weight of 0 is set when the difference is 5°C or more. The values between 0°C and 5°C are scaled proportionally so that the weight of 0.5 corresponds to the temperature difference of 2.5°C. The freezing point has no temperature difference attribute similar to the dew and frost points. Instead, we had to calculate the difference ourselves. The same values were used for the weight generation as for the dew and frost point differences. Additionally, there are two freezing points, one representing the "true" freezing point while the other one has an additional safety buffer added to it. As the true freezing point is more common, we elected to use it over the "safety point," the 5°C gradually increases weight as the freezing point approaches, providing a buffer.

\subsection{Visibility}
Poor visibility increases the risk of accidents as drivers struggle to observe their surroundings. The Digitraffic API provides visibility as both meters and kilometers, produced by two sensors. The DSC and the PWD12 sensors can detect distances up to 2 km while the PWD22 sensor can detect them up to 20 km \cite{RWS200ProductCatalog}. These two distinct max values make determining the weight scale a challenge; however, based on data queries made on clear days, it would seem that the majority of stations have equipped the PWD22 sensor and can detect distances up to 20 km. The PWD values capped at 2000 m seem to appear only at 13 stations along Road 1 in southern Finland, one station in Keuruu, and four stations located at airports. Therefore, we will ignore these outliers and consider 20 km as the maximum detectable value. Visibility can be considered good when it is more than 10 kilometers, moderate when 4-10 kilometers, poor at 1-4 kilometers, and very poor under 1 kilometer. Visibility below one kilometer indicated the presence of fog \cite{ilmatieteenlaitosNakyvyys}. Based on this, the weight of the visibility factor was set to 0 if the detected visibility was 10 km or more and 1 should the visibility have fallen to 0. The values in between were scaled proportionally.


Fog can form when the dew point falls below the temperature of the air. The API contains attributes for the dew point and the difference between it and air temperature. For assessing driving conditions, the difference attribute gives direct value, as a more negligible difference would mean a higher risk for fog formation. The frost point was considered in temperatures below freezing, similar to what was discussed above with the road surface weights. The value range considered here is the same as above, i.e., the weight of 0 for a difference in temperatures of 5°C or more and 1 for differences of 0°C and less. The relative humidity of air is also closely related to this. Fog tends to form in conditions when the relative humidity is at least 90\%. Therefore, the closer the percentage from the API is to 100, the closer the cost is to 1.


Precipitation of various types is another major factor affecting visibility. The type of precipitation, as well as its intensity, plays a key role in determining the effect it has on driving conditions. Rain, sleet, and snow are each associated with a higher risk of accidents than the previous, which also increases with the intensity of the precipitation \cite{malin2019accident}. The Digitraffic API has multiple attributes detailing the precipitation conditions on the road. The system can determine if the precipitation is drizzle, rain, snow, wet sleet, regular sleet, hail, ice crystals, snow grains, snow pellets, or freezing drizzle or freezing rain. In some cases, when the precipitation is very weak, it might not be possible to determine the type. 

Both Tobin et al. \cite{tobin2021adverseweather}
and Malin et al. \cite{malin2019accident} found snow more perilous than rain. Our ranking of the precipitation types was primarily based on these two previous works, as most of it focuses on precipitation intensity or uses even more crude categorizations of precipitation, such as only considering regular rain versus snow \cite{andrey2003weather}. According to Malin et al., sleet (rain/snow mixed) is more dangerous than rain but less dangerous than snow. Freezing drizzle and freezing rain were not ranked with the other precipitation types, but the relative risk associated with them would fall between sleet and snow. In comparison, according to Tobin et al., snow is associated with a higher risk than rain but a lower risk than freezing rain. Sleet (ice pellets) is considered more dangerous than rain, but defining a more specific spot for it in the hierarchy was not possible. Any precipitation type is linked to a higher relative risk than dry weather. 

Dry weather was considered the optimal conditions, therefore it was assigned the weight of 0.0. Precipitation that was too weak to determine its type was considered the second most optimal, followed by the various types of rain ranked based on their intensities. Neither paper distinguished between wet and regular sleet (rain/snow mix), but we reasoned that since wet sleet is closer to regular rain, it would be considered less dangerous than regular sleet and could be considered more similar to snow. Hail was excluded from analysis by Tobin et al.\cite{tobin2021adverseweather}, nor could comparable studies that would have specified the risk associated with hail. We reasoned that they should be considered more of a hindrance than sleet as hail tends to fall rapidly in large quantities, impacting visibility and potentially causing loss of traction, but less troublesome than snowfall as hailstorms are typically shorter in duration. 
Ice crystals, snow grains, and snow pellets were considered sub-types of snow. The final challenge was deciding which should be ranked more dangerous: freezing rain or snow. It is possible that geographical distance explains the difference in accident risks in a study done on Finnish data \cite{malin2019accident} and a study done on data from Kansas \cite{tobin2021adverseweather}, as the intensity and frequency of the phenomena can be vastly different. Legislative differences may also play a part; for example, winter tires are mandatory in Finland but are not necessarily so in the US. As a compromise, we ranked freezing drizzle below the snow in danger, while freezing rain was placed above the snow. 



If the detailed precipitation type is unavailable, we elected to use another attribute, classifying precipitation as either rain or sleet/snow. To emulate the weights from the previous attribute, if the reported precipitation type is sleet/snow, the weight will be set to 0.722, the average for sleet and snow, as we cannot know which type is in question at any given time. The same attribute also classifies precipitation based on intensity as weak, moderate, or abundant. Again, to best emulate the weights determined with the more precise attribute, weak rain was likened to drizzle and given the weight of 0.222, while moderate and abundant rain was given the weight of 0.333.

Precipitation status was also provided in two more attributes, first of which provides a simple boolean of whether the station is currently perceiving precipitation, and the second provides a more varied enumerate value. These two were not used as the boolean attribute is present at very few stations, while the meta did not provide value descriptions for enumerate attribute, making deciphering the values next to impossible.

The rain intensity is also provided in milliliters. The Swagger documentation of the API classifies milliliter rain intensities into similar intensity categories with slightly different names, such as light rain, rain, and heavy rain. Rain is considered heavy when the intensity is more than 7.6 mm per hour. Rainfall of 7 mm/h or more is considered heavy rain by the Finnish Meteorological Institute. To account for differences even with more intense rainfalls, we set the value at the maximum weight of 1 to be 10.0 mm/h. Unsurprisingly, the minimum weight of 0 was set to 0.0 mm/h, with the values in between scaling proportionally.
Additionally, the system records the sum of daily precipitation. These values are not too relevant to us, as their primary importance would be to assess how much water could be on the road's surface. For this purpose, we already have another attribute measuring this directly. 


\subsection{Other Environmental Factors}
While the influence of visibility and road surface conditions on driving is rather obvious, other environmental factors are more easily overlooked. High winds can make steering challenging and pick up dirt that decreases visibility. They can also cause trees to fall on the road, causing obstructions. Similarly, both extreme cold and heat can hinder travel. Among other consequences, cold weather can cause condensation in the fuel tank, causing engine problems, while hot weather can lead to the engine overheating. 

The API offers data on the average and maximum wind speeds from the past 10 minutes, in addition to providing the average direction of the wind.
While the wind speeds can be used for weight vectors independently, the direction is different. To make it relevant for driving, one would need to know the direction in which the vehicle would be heading on the road. This would increase the weight on segments experiencing crosswinds, which can be particularly dangerous. However, this process was deemed outside the scope of this project, and therefore, only the immediately usable attributes of average and maximum wind were used. The highest possible wind speed reported by the RWS200 weather station is 75 m/s \cite{RWS200ProductCatalog}. The Finnish Meteorological Institute (FMI) categorizes winds into seven categories based on average speeds from 0 m/s to 33 m/s and faster, defined as hurricane \cite{ilmatieteenlaitosTuuli}. However, such a high wind average has never been measured in Finland. 
Winds measuring 21 m/s and faster are classified storm winds \cite{ilmatieteenlaitosTuuli}, leading up to which the FMI issues wind warnings once the average wind speed reaches 15 m/s (in summer) or 20 m/s (in winter). 
It is normal for wind gusts to be 1.5 to 2 times more intense than the measured average winds. Therefore, it was justified to consider different weights for the average and maximum winds. Based on the previous, we set the maximum average wind to 21 m/s, which would be given the weight of 1, while the maximum value for maximum wind speed was set to 31.5 m/s, which is 1.5 times the maximum average speed we set. Winds of 0 m/s were set as the minimum for both.

As previously stated, extreme temperatures can have a detrimental effect on driving conditions. The Digitraffic API provides multiple temperature measurements on top of the road surface temperatures discussed earlier. These include current air and ground temperature, as well as the minimum and maximum air temperatures from the past 24 hours. For now, the historical values were discarded as the focus is on the current conditions. We also ignore the ground temperature, as we believe that the temperatures from the air and the road surface are more pertinent to driving conditions.
The FMI releases an alert for heat once the temperature is predicted to reach +27°C. 
We will consider this our max heat value and assign any vector with this value or higher the weight of one. Defining the threshold for harsh frost is more complex, as the FMI has determined different thresholds for different parts of the country. The Northern Ostrobothnia falls into the northern part of the country, which would place our other max value at -30°C. The midpoint between the values would be 0°C, and the weight could be determined by the difference between the midpoint and the current air temperature, with 0.0 degrees leading to the weight of zero and the difference of 30 degrees to the weight of one. According to the FMI, the "optimal" temperature in Finland is 14°C, with rising temperatures bringing adverse consequences much faster than a drop in temperature does. 
In this case, the temperature of 14°C would yield the cost of 0, -30°C and +27°C the cost of 1, and values in between would be weighted based on the relative distance from 14°C. For example, the temperature of 20.5°C would be assigned the weight of 0.5 as it is midway between the optimal temperature of 14°C and +27°C. The weight of 0.5 would also be assigned should the air temperature be -8°C as this is the midway between the optimal temperature and -30°C.

\subsection{Vehicle Volume}
There are 520 traffic measurements in Finland, 15 of which have been temporarily removed, leaving 505 stations actively gathering data. The stations use a pair of induction loops to detect the passing vehicles, their lengths, and their speeds. The stations can also distinguish between directions and lanes used by the vehicles. Traffic measurement stations produce fewer attributes than weather stations do, and they also come with additional documentation, leaving less room for misinterpretations. The meta introduces 103 traffic measurement attributes, numbered from 5016 to 100000. As with the weather attributes, many attributes can be considered sisters of one another, for example there are multiple attributes describing the travel speeds from the past 5 minutes on different lanes. The attributes can be categorized into three distinct groups: those detailing the speed of vehicles, those focused on the number of vehicles passing the station, and attributes describing the dynamic and static characteristics of the station itself. While the vehicle volume and speeds are monitored continuously, depending on the attribute, the corresponding values are not updated quite as often. Update intervals can range from three to five minutes up to an hour. There are also differences between stations: some boast a continuous real-time update rate, while most stations have an update interval of five minutes. The data items used in the weight creation can be seen in Table \ref{tab:weights}. 

The Highway Capacity Manual (HCM) offers a guide for categorizing roads based on the current Level of Service they can offer under the present circumstances. The LOS is a scale from A to F, with A being the best possible rating.
Multiple things affect the LOS category of a road, including vehicle volume and speeds \cite{hcm2010}.
When the volume of vehicles on the road increases beyond the maximum capacity, travel speeds start declining, congestion can form, and the risk of accidents increases. In the HCM, whether the vehicle volume exceeds the road capacity is used as a final threshold for determining whether the LOS of the road is classified as an E and an F \cite{hcm2010}. In route planning, one should strive not only to avoid roads experiencing congestion but also those where the risk of congestion is high. Therefore, instead of only concentrating on whether the volume exceeds the capacity, we focus more on how close the volume is to exceeding capacity.
The maximum capacities of each road segment are presented in the API, therefore we need not calculate the capacities ourselves.

The attributes come in two update types: the first one updates the data continuously according to the update rate of the station in question to reflect the situation within the past 3 or 5 minutes, while attributes of the other type are updated in regular 5 or 60-minute intervals. All values are extrapolated to one hour if that is not already the time window used. Another divisive factor would be the level of specificity of the attribute: some attributes describe the volume on a specific lane, others per direction, and one even overall.
There are huge differences in the availability of the attributes: while there are about a hundred stations that offer lane-specific data about volumes, this is a clear minority. For the vast majority of the stations, the maximum level of detail is the volume per direction. Additionally, many roads in Finland only have two lanes, one in each direction, rendering lane-specific details redundant. There is also really no sure way of attributing each of the fourteen possible attribute lanes to a physical lane based on the limited information that we have. Therefore, we discarded all lane-specific attributes. 
We also discarded the attribute that did not differentiate between directions at all.

Multiple attributes can be used to monitor the number of heavy goods vehicles on the road. This data could be used to identify routes frequented by the vehicles, which could be useful for determining personalized routes for users who dislike driving among heavy goods vehicles. The data could also be experimented on to track exempt transports that utilize large vehicles. However, this sort of analysis is beyond the scope of this paper, therefore the data pertaining to the heavy goods vehicles is discarded at this point.


After discarding the lane-specific attributes, attributes focused on heavy goods vehicles, and the attribute combining all vehicles passing the station, we are left with 10 direction-specific attributes. While some stations might be located at crossroads or on bridges, they still only monitor the traffic on one specified continuous road, meaning that there are only ever two directions. 
In the traffic measurement data API, each direction has its own attributes. Half of these attributes report the number of vehicles passing the stations per hour, while the other half gives the percentage of how much the volume is of the maximum capacity of the road. Since the capacities differ greatly between roads, the flat volume does not communicate much about the conditions on the road. In contrast, the volume-to-capacity percentage is more telling as it takes into consideration the characteristics of each road, such as the number of lanes. By discarding all flat values,
we are left with six attributes, three per direction. The main difference between these attributes is their update cycle. The first one is updated continuously to reflect the conditions experienced in the past five minutes, while the other two are updated at regular intervals, either five or 60 minutes. 
Because situations on the road can change quickly, we wanted to use an attribute as close to real-time as possible. Therefore, we elected to utilize the continuous volume-to-capacity ratio as our attribute for vehicle volume.

For route planning, we do not only wish to avoid routes that are already congested but also those that have a high risk of becoming congested in the near future. The smaller the percentage is, the smaller this risk is. The higher, the bigger the risk is. By this logic, when the attribute's value is 0\%, the appropriate road segments will be given the weight of 0. Similarly, if the value is 100\% or more, the weight will be 1. Values in between are proportional to the percentages, e.g., with the volume reaching 50\% capacity, the segment's weight will be 0.5.

\subsection{Vehicle Speed, or Free Flow Speed (FFS)}
Tracking vehicle speeds is important as it provides the means for detecting obstructions on the road that might slow down traffic flow. These obstructions can be traffic jams but might also include slow-moving maintenance vehicles or traffic accidents. Slowing speeds can indicate issues even before the vehicles start to pack up, in other words, before the volume increases. The Digitraffic attributes for vehicle speeds also contain those focused on specific lanes and directions, just like volume attributes do. As with volume attributes, we disregard the lane-specific speed attributes.
There are no attributes dedicated to heavy goods vehicles like there were with volume attributes. Instead, there are attributes dedicated to the slowest detected speeds. These could be useful for detecting abnormalities in the traffic flow, but for now, we discard these attributes as their relevance to traffic conditions can be considered trivial.

The attributes 10 left are very similar to the volume attributes in that six attributes only communicate the flat speed value measured within a 5 or 60-minute window, and six attributes are updated in regular intervals while four are updated continuously. Eliminating the attributes providing flat speed values,
as well as those updated at regular intervals instead of continuously
left us with two attributes: the average speed in the past 5 minutes in either direction as a percentage of the free flow speed (FFS).
The directions associated with these attributes are the same as those concerning the vehicle volumes, therefore as we had already determined the directions to note for each station with the volume attributes, we could use these results for the speed attributes as well.

The FFSs of each station are defined in the station meta and the same list of constants that contained the capacity definitions. Each direction can have a different FFS. Additionally, on some roads, the FFS is different depending on the time of the year. Digitraffic also provides classifications for the percentages to increase human readability. When the FFS is 0 to 10 percent of the FFS, it can be said that the traffic is stationary, at 10 to 25 percent the traffic is queuing, 25 to 75 the traffic is slow, 90 to 100 the traffic is fluent. If the percentage is larger than 100, the current speed is higher than the FFS \cite{lamDocumentation}. A lower percentage signifies degraded traffic conditions, and the smaller the attribute value, the higher the cost of the related road segment. Therefore, if the value of the average speed relative to the FFS is 0 \%, then the weight of that road segment is 1. If the relative speed is 100 \%, the weight is 0. Values are between are scaled so that 50\% is 0.5, 75\% is 0.25, etc.

\subsection{Traffic Alerts}

Digitraffic relays traffic alerts related to four primary topics:  exempt transports, weight restrictions, maintenance, and traffic announcements, which can include things like accidents.
The traffic announcements can be retrieved in either JSON or DATEX II format through the API. When received through the MQTT WebSocket API, the announcements are gzipped and must be decompressed before they are readable.

\subsubsection{Road Work}
The road work API provides information about current and upcoming road work operations. The descriptions can include work times, start and end times, working hours, work types, and the disturbance caused by traffic. This description includes the type of restriction imposed on traffic (closed lane, lowered speed limit, detours, etc.), whether the restrictions can be lifted, and the severity of the disturbance (Low, High, Highest). In route planning, the current road work must be retrieved through the JSON API, as the road works are only published through the WebSocket when they are new or updated. Additionally, many road work alerts contain information about road work planned out into the future. This means that utilizing the road work data requires more processing than, for example, weather data, which is always current. A small hindrance to using this data is that it is clearly made to be very human-readable, with less consideration for machine readability. This is especially apparent with the free input comments and road work type descriptions, which appear to be manually written based on variations in capitalization, use of near-synonyms, and mentions of specific roads or buildings within the description. 

Different types of road work would generally warrant different types of mitigation procedures for route planning; however, as the type of road work can be almost anything thanks to free-form typing, this is not possible. Therefore, our focus was on the severity of the disturbance, and weights to vectors representing the affected road segments will be allocated according to it. If there is no road work affecting the segment, the weight is defined as 0. If the severity is defined as low, the weight is 0.33, if High the weight is 0.66 and finally, should the severity reach the rating of "highest", the cost is set at 1. 


\subsubsection{Traffic Announcements}
Traffic announcements are arguably the most relevant category of the four, these types of alerts provide information about events that have an immediate affect on traffic. Accidents are one of the most common announcements topics provided in this category. Other topics include vehicles broken down on roads, animals on the road, large fires that affect traffic, disturbances to ferry schedules, and more. As there is no definitive list of all of the possible types of alerts and no list for all types of road work, we could not devise situation-specific procedures for route planning. To add a layer of complexity, alerts come in six set types: general, retracted, unconfirmed observation, ended, and the accident-specific preliminary accident report and accident reports. 

We focused on handling accidents due to lacking a definitive list of topics. Typically, once an accident occurs, the first alert will be a preliminary accident report, followed by an accident report or a general alert if the situation changes or more details become available. Once the accident has been cleared, an "all clear" is released. However, sometimes, the only alert that gets released about an accident is the preliminary report. In such situations, it can be difficult to know when the route is sensible to travel again, as there is very limited information about the scale or effect of the accident. From their own web UI, Fintraffic removes preliminary accident reports without further details after about 30 minutes.
As there is no guarantee of detailed information, we assumed that all accidents are detrimental enough to the traffic flow that the road should be avoided. Therefore, any road segment with an accident is given a weight of 1. If the road segment has no accident, the weight is set to 0. A segment would be considered to have an accident if a) a preliminary accident report was published in the past 30 minutes or b) an accident report or general announcement about an accident has been released with no announcement of the situation being over.

\section{Road-weather Aware Routing}

\subsection{Routing Algorithm} 


\subsubsection{Data extraction for graph creation}

\begin{figure}[H]
    \centering
    \includegraphics[width=0.8\linewidth]{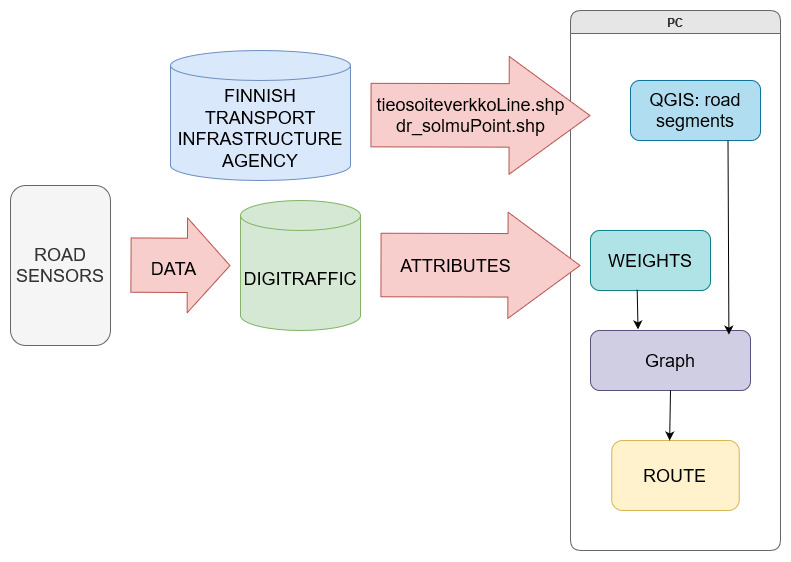}
    \caption{An overview of the different stages of the route creation.}
    \label{fig:route_overview}
\end{figure}

As shown in Fig. \ref{fig:route_overview}, the first stage of the routing algorithm is the data extraction stage. The data from road sensors is collected by Digitraffic, where attributes are then extracted to form the weights for the algorithm. The road data was collected for road segmentation so that a graph with edges and nodes could be created for the algorithm. The road information, as shown in Fig. \ref{fig:route_overview},  was downloaded from Väylävirasto's website Suomen Väylät \footnote{https://suomenvaylat.vayla.fi/?lang=en}. Two ShapeFiles tieosoiteverkkoLine.shp and dr\_solmuPoint.shp were downloaded. tieosoiteverkkoLine.shp file contained road addresses, road numbers, road section numbers, carriageway numbers, and distance from the start of the road section. The file also contained descriptions of the road connections to other roads. According to Väylävirasto, locating the correct position is possible within 1m accuracy based on this. The road segments in Väylävirasto material have nodes at the beginning and end of each segment, and dr\_solmuPoint.shp file contains these nodes \cite{digiroad2024}. It includes each node identification number, coordinates, and other information, such as municipality code. 

The tieosoiteverkkoLine.shp and dr\_solmuPoint.shp files were then further processed in QGIS with PyQGIS. This was done so that the roads would be divided into road segments at every intersection of two or more roads. Tools native:geometrybyexpression and native:splitwithlines were run with processing. The geometrybyexpression tool was run on the nodes from dr\_solmuPoint.shp file so that small lines were created on the node locations. Then native:splitwithlines was run with the small lines file and tieosoiteverkkoLine.shp, and it found the intersections of the lines in those files and created a new file with the road segments split at the node locations. This meant that for example the road segments did not end up going over crossroads, which would have made it impossible for the algorithm to take turns at intersections. 

\subsubsection{Graph creation}
The graph for processing the routes was created using NetworkX's Graph tools. The file used as a basis for this was the road segment file created in the previous step. Each road segment was read from the file one at a time so that the coordinates of the end and start nodes were read, as well as a weight vector for the road segment. The weight vector creation is described later. These were saved into the graph by taking each road segment and adding it as an edge. 

\subsubsection{Route creation} \label{routecreation}
The route itself is created using Dijkstra's algorithm from the NetworkX library. The start and end nodes were given as attributes, as well as the graph and a weight function. The end and start nodes were given as coordinates, and they were confirmed to be included in the nodes of the graph before route creation. If they could not be directly found, the closest node to the coordinates was searched and used as a start or end node. The weight function is described next. Dijkstra's was chosen as it is simple algorithm and well-known, and it allowed us to focus more on the attribute use, what data attributes to choose for personalization use, and how well they functioned. Our future work will compare the Dijkstra's to Graph Neural Network (GNN) and other possible algorithms. GNN would be more complex, possibly with higher computing overhead, but it would allow the algorithm to learn the driver's preferences.

\subsubsection{Weight function} \label{weightfunction}
The weight function was a dot product of the weight vector, which contained the data attributes and a preference vector. The preference vector was created for different example users. These are described better in Section \ref{sec:eval}. Dot product was used because Dijkstra's required one single value for each edge to be used as a weight. The preference vector is where the personalization happens, as it allows us to estimate how the user would prefer certain attributes to be and, eventually, what kind of routes they prefer. As the weight function is a single value, it equalizes the influence of different attributes. It is thus important to get the preference vector right, as it has a preference value for each attribute. After the dot product calculation, the attributes are included in the same number. The graph and route are then drawn using NetworkX's drawing tools. 

\subsection{Data Fetching Service}
A primitive MQTT listener was built in JavaScript to collect weather and traffic data, as well as current traffic announcements, for a proof-of-concept. While the listener runs, it continuously updates the array associated with each selected weather and traffic station with the newest data. Traffic alerts in North Ostrobothnia are also saved into the browser's memory, as well as the data from the road stations when receiving the alert, which could then be downloaded.
We used Postman to fetch all recently updated road work instances filtered manually to attain the road work sites in our area of interest, as the API does not allow for conditional queries. For each road section, we only considered one road work site, even in cases where there were several. We also neglected the working hours and considered all work sites as active, even if this did not represent reality to the highest possible degree. It was reasoned that even if work is not actively being conducted, abnormal traffic arrangements often persist outside of working hours.

\subsection{Experimental Setup}
In our work, we focused on the southwestern part of the Northern Ostrobothnia region, excluding the island municipality of Hailuoto. The northern border of the area was set to follow the river Kiiminki, in the west the border falls along the coastline, in the south along the border of the region, and in the east either the regional border or the border of the excluded municipalities of Raahe or Utajärvi. From the measurement stations in this area, we removed some stations such as four test stations, 
are stations marked as "removed temporarily,"
and one station that produced no data despite the meta claiming otherwise.
In two instances, we combined two stations located less than a hundred meters from one another into one station. 
In total, we had 27 weather stations and 37 traffic measurement station locations in our area of interest. These stations employ a variety of sensors that sometimes differ from one another. Some stations give a much more detailed view of the conditions, while others give more of an overview. This is concretely visualized by examining the number of attributes from each station: the weather and traffic measurement stations with the highest number of attributes produce 120 and 30 attributes, respectively, while the stations with the smallest number of attributes produce only 23 and 8, respectively.

Within this area of interest, we decided to focus on a trip from Oulu (more specifically, the University of Oulu) to Ylivieska, a town some 120 km to the southwest. Each road segment was assigned both a road weather station and a road traffic station. Typically, the geographically closest station was selected. In certain instances, however, manual adjustments were made to prioritize stations located on the same road as the segment over those situated on adjacent roads.

\begin{figure}
    \centering
    \includegraphics[width=0.8\linewidth]{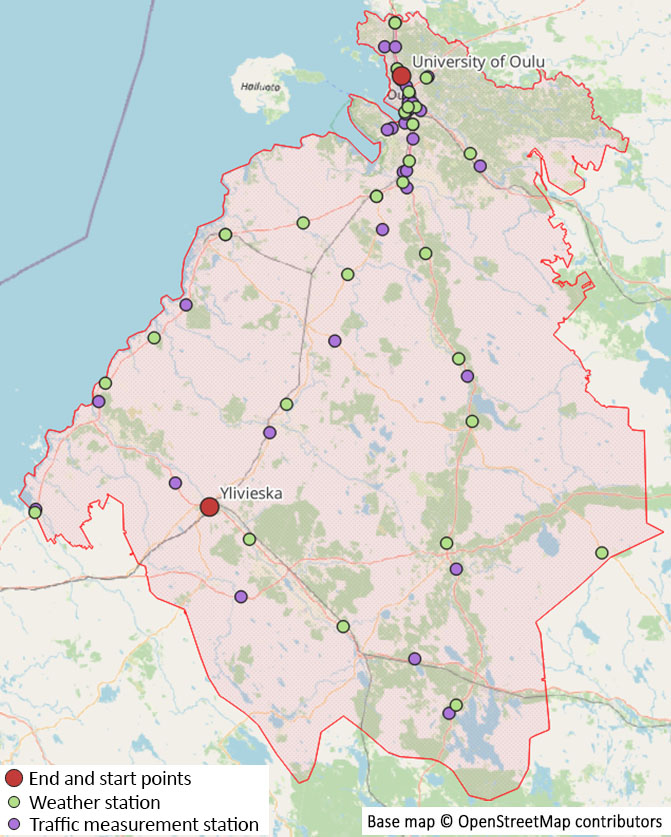}
    \caption{The area of interest outlined on a map. Lavender dots denote the location of traffic measurement stations, while the green dots signify weather stations. The locations of the University of Oulu, as well as the city of Ylivieska, are marked with red, slightly larger dots. Please note the distance disparity between the stations near the city of Oulu and the area further to the south.}
    \label{fig:areaofinterest}
\end{figure}

The MQTT data fetching service was used to gather data on different situations on the roads. Three such situations were used in this experiment: a late summer thunderstorm, a moderately slippery October morning, and a seemingly uneventful afternoon to set what could be considered a baseline.
The data from each measurement station was pre-processed so that in cases where there were duplicate attributes, the average of those present was calculated, and if a certain attribute was missing, whenever possible, it was supplemented with a similar attribute (as described in Section \ref{roaddata}). 
While screening the data, we noticed a strange pattern at one of the road traffic stations, wherein it seemed to always experience moderate congestion. Upon further investigation, it was determined that the FFS defined in the meta for this specific station was most likely set at a time when the speed limit was higher – the FFSs are defined as 80 and 78 km/h for either direction, while according to visual inspection on site, and the speed limit data from Väylävirasto sets the speed limit at 60 km/h. 
To correct any skewing this might cause, we always calculated the FFS of the station in the program based on the FFS of 60 and 58 km/h, following the example of the values in the meta but lowering them to accommodate the lower speed limit. The percentage was also calculated in the program whenever the ratio variable was not available. However, with other stations, the FFSs defined in the meta were used instead, as no similar abnormalities were noted. 

The baseline dataset was acquired around noon on a calm Tuesday in August. Both weather and traffic weights calculated on this dataset see only small differences between stations. The weight ranges at weather and traffic stations were [0.2425, 0.0543] and [0.0, 0.3], respectively.
The two other datasets saw more variety between stations. Late summer thunderstorms are often highly localized and can be accompanied by exceptionally heavy rainfall. This was visible in the second dataset, which was collected similarly at the end of August but under significantly different weather conditions. On a thunderous day, the weight ranges at the weather and traffic stations were [0.1373, 0.5845] and [0.0, 0.52], respectively. On a slippery October morning, the ranges were similar to the thunderstorm, with weather weights varying between [0.1611, 0.5404] and traffic weights between [0.01, 0.525].

To achieve these weights, the collected data was processed as follows. The data from the poller was saved into text files. The text files contained dictionaries of each type of measurement station, as well as arrays containing the sensor-value pairs that each station had produced. The sensor-value pairs could then be extracted in the route planning program to serve as input for the weight creation for each station. If there were no data from a certain station, that station would be assigned a secondary station, from which data was used instead. The assignment was done manually and only in cases where there was absolutely no data produced by the station. 

The weather data was used to create three weather weight factors for each station, namely the surface, visibility, and other environmental factors. To do this, each relevant attribute was assessed as described in Section \ref{roaddata} by taking the average of the values of duplicate attributes (whenever present) and calculating the factor value by linearly scaling the attribute value between the minimum and maximum values defined in Section \ref{roaddata}. The average of these factor attributes was then used to determine the corresponding weight factor. 
These weight factors were then used to determine two weather station weights: full weather, which contained the surface, visibility, and other environmental weather, and environmental weather, which comprised the latter two. If a secondary weather station had to be used, only the environmental weight was used as the surface factors were determined.

\begin{figure}[H]
    \centering
    \includegraphics[width=1\linewidth]{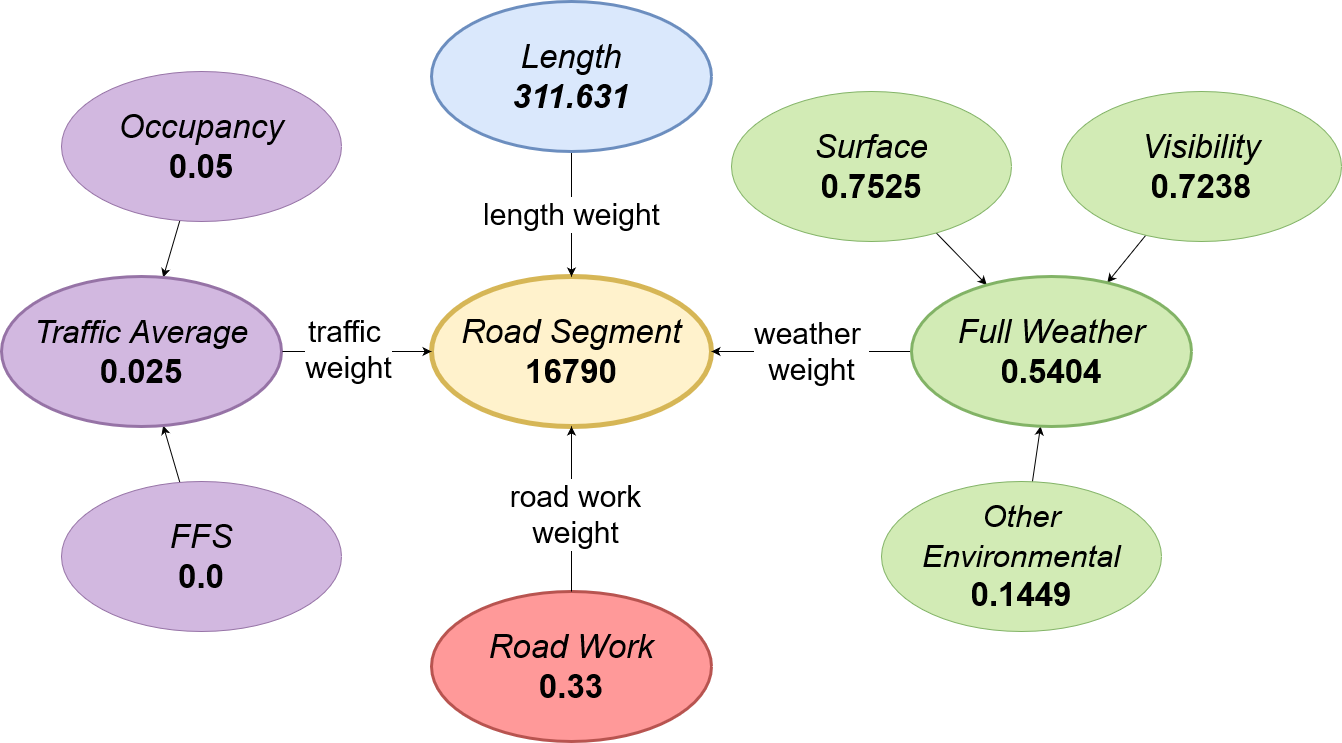}
    \caption{Example of weights calculated for a single road segment.}
    \label{fig:examplesegment}
\end{figure}

The traffic weight calculation was done similarly. The factors considered were the FFS and the occupancy of the road, and the final traffic weight was determined by calculating the average of these factors. Once the weights had been calculated for each station, these weights could then be assigned to the road segments to be used in the route planning. Possible maintenance events were also assigned to the road segments at this stage. The weights affecting a single road segment have been depicted in Figure \ref{fig:examplesegment}. The final route weights are determined based on the user-specific preference vectors, as described in Section \ref{routecreation}.

\subsection{Evaluation with User Stories} \label{sec:eval}
We created user stories for three distinct kinds of users. "Teemu" is a 22-year-old man who has a brand-new sports car. Teemu likes to drive fast, and enjoys it immensely, therefore, he is not too concerned by the length of the trip. Much more so, he prefers straight roads with as little traffic as possible. He also prefers avoiding road work sites but is not too bothered by them. Teemu is adamant about avoiding road salt to prevent his new car from rusting.
"Tapio" is a 45-year-old man who works in Ylivieska. Tapio is a very seasoned driver, and as such, he is used to driving in all kinds of weather. He does not care too much about traffic or roadwork, and he is only concerned about finding the shortest and fastest route to Ylivieska.
"Tuire" is 72 years old; she lives in Oulu but often visits her grandchildren in Ylivieska. Due to her advanced age, her reaction time has increased, and as such, she tends to avoid routes with anything out of the ordinary, such as road work sites and adverse weather conditions. On the other hand, congestion and the distance itself are not that important to her. 

Based on the user stories, we evaluated how important the user would perceive the length of the route, the amount of traffic, weather conditions, and the presence of the road work sites. Traffic, weather, and road work sites were chosen to reflect each of the main datatypes present in the datasets. Length was included as it is the traditional measure of the shortest path. Each of the four factors was rated either unimportant, somewhat important, important, or very important. Based on the ratings, preference vectors were created for the users so that unimportant factors were weighted at 0.05, somewhat important factors at 0.25, important ones at 0.5, and very important factors were weighted at 0.75.
As such, the initial preference vector of Teemu was set at [0.05, 0.75, 0.25, 0.75] for length, traffic, weather, and road work sites, Tapio's at [0.75, 0.25, 0.05, 0.25], and Tuire's at [0.05, 0.05, 0.75, 0.75]. These vectors were then scaled so that their sum is 1.

The scaling was done by taking the sum of the weights and dividing the weight of each individual factor (length, traffic, etc.) by that sum so that all preference weights of each user would add up to 1. The final preference vectors could then be determined. The preference vector for Teemu, which prioritizes minimum traffic and road works with some consideration for the weather, was set to [0.03125, 0.46875, 0.03125, 0.46875] after scaling. Tapio, whose main priority is to follow the shortest path while weather is mainly ignored, was given the preference vector [0.576923077, 0.192307692, 0.038461538, 0.192307692]. Finally, Tuire, who was first and foremost concerned with avoiding adverse weather and road work sites but did not mind the distance or traffic, was designated the scaled preference vector of [0.03125, 0.03125, 0.46875, 0.46875].
An overview of the vectors is shown in Table \ref{tab:preference-table}. 
These vectors were then used in the weight function as described in section \ref{weightfunction}.

\begin{table}[]
    \centering
    \scriptsize
    \begin{tabular}{|c|c|c|c|c|} \hline
        \textbf{User} & \textbf{Distance} & \textbf{Traffic} & \textbf{Weather} & \textbf{Road Work} \\ \hline 

        Teemu & 0.03125 & 0.46875 & 0.03125 & 0.46875 \\ \hline
        Tapio & 0.576923077 & 0.192307692 & 0.038461538 & 0.192307692 \\ \hline
        Tuire & 0.03125 & 0.03125 & 0.46875 & 0.46875 \\ \hline
         
    \end{tabular}
    \caption{The three preference vectors created based on importance of each factor derived from the user stories. }
    \label{tab:preference-table}
\end{table}


\section{Results}

\begin{figure}
    \centering
    \includegraphics[width=0.8\linewidth]{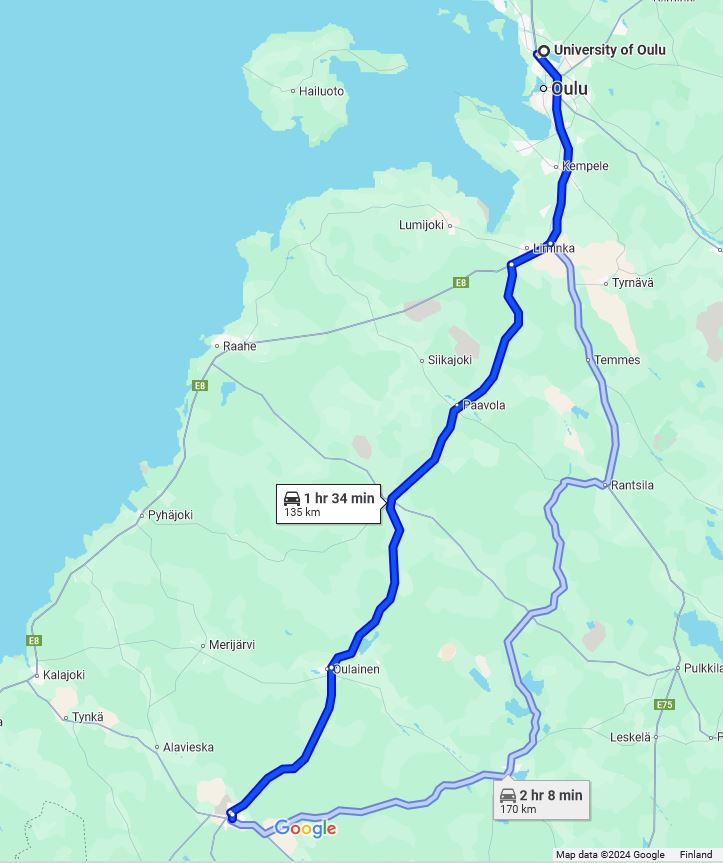}
    \caption{Routes suggested by Google Maps.}
    \label{fig:mapsroutes}
\end{figure}

\begin{figure}
    \centering
    \includegraphics[width=1\linewidth]{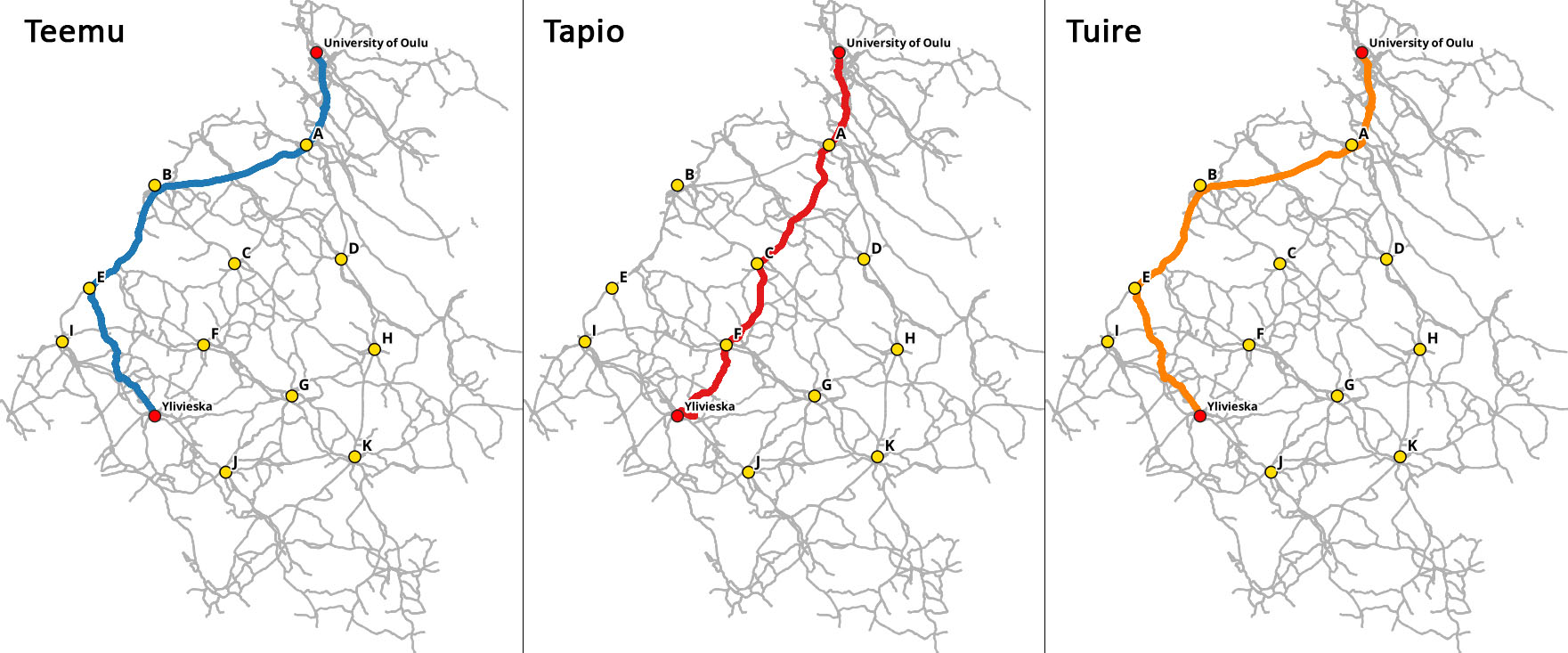}
    \caption{Routes in the base dataset.}
    \label{fig:baseroutes}
\end{figure}

\begin{figure}
    \centering
    \includegraphics[width=1\linewidth]{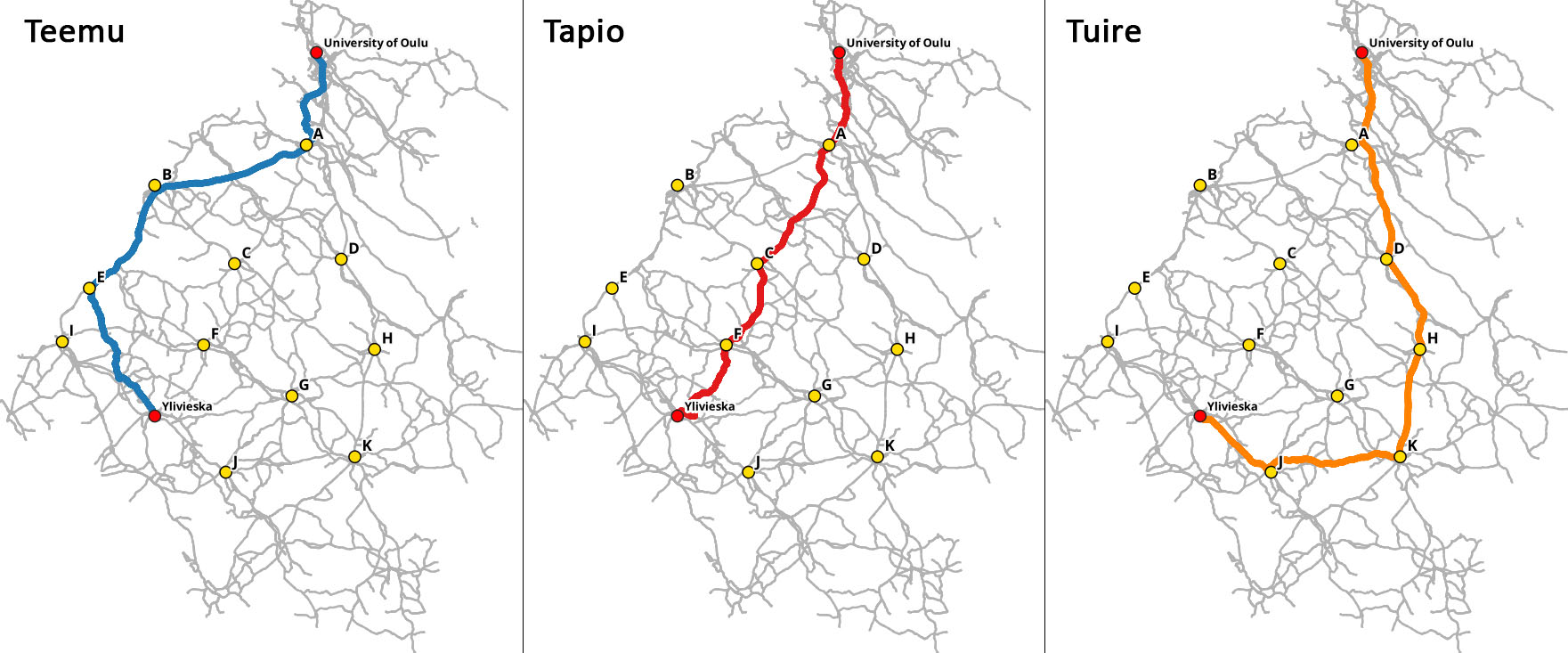}
    \caption{Routes in the thunder dataset.}
    \label{fig:thunderroutes}
\end{figure}

\begin{figure}
    \centering
    \includegraphics[width=1\linewidth]{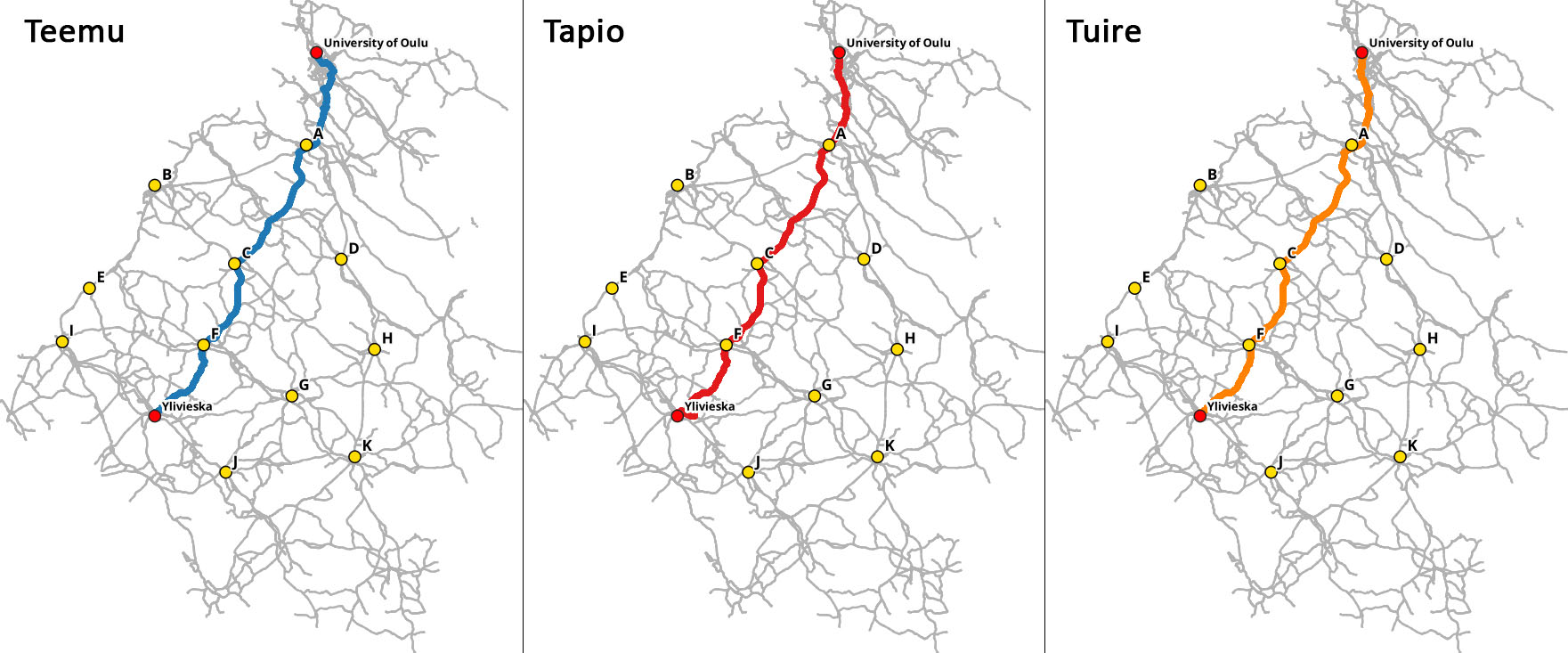}
    \caption{Routes in the early winter dataset.}
    \label{fig:octoberroutes}
\end{figure}

We created routes for each user with each dataset, nine routes in total. Figure \ref{fig:baseroutes} presents the routes created for the base data, Figure \ref{fig:thunderroutes} shows the routes for the thunder data, and finally, Figure \ref{fig:octoberroutes} presents the routes for the early winter data. Notable crossroads are marked with yellow circles and assigned a letter A-K, while the start point at the University of Oulu and the finish point in Ylivieska are marked with red circles.
Tapio, who only cares about getting to Ylivieska as soon as possible, is given the same route each time, from the University through points A-C-E to Ylivieska. The routes calculated for Tapio most closely resemble the suggested by Google Maps presented in Figure \ref{fig:mapsroutes}. Teemu, who is first and foremost concerned with traffic and road work sites, was primarily routed around those. As such, the first two of his routes were mostly the same, from the University via points A-B-E to Ylivieska, with a small difference at the very beginning, while in the third data set, his route was similar to Tapio's. Tuire had the most different routes for each situation, as her routes avoided road work sites and adverse weather. In both the base dataset and early winter dataset, her route closely follows that of Teemu's, while in the thunder dataset, her route diverts inland to avoid the thunderstorm closer to the cast. The route goes from the University through points A-D-H-K-I to Ylivieska.

\begin{figure}
    \centering
    \includegraphics[width=1\linewidth]{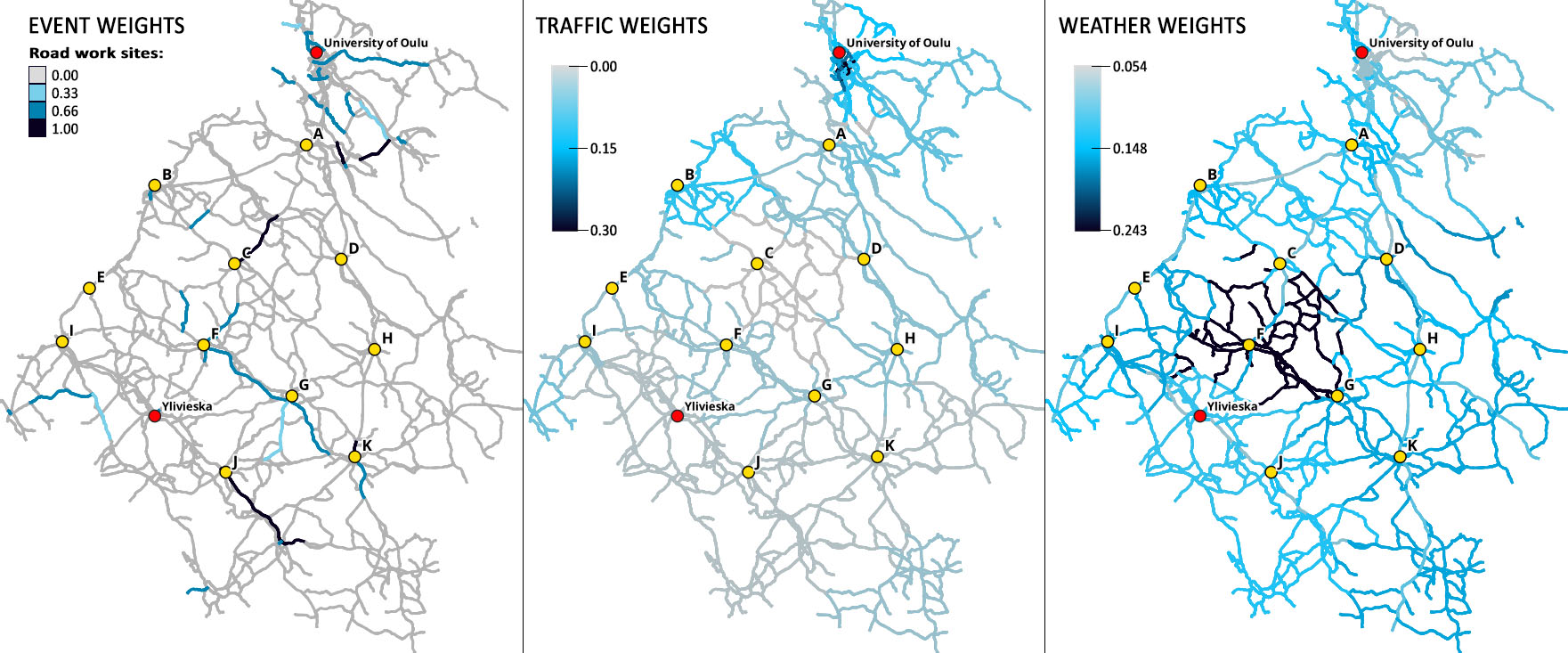}
    \caption{Weight distribution in the base dataset for events, traffic measurements, and weather in the area of interest.}
    \label{fig:basemap}
\end{figure}

Despite the dataset lacking any adverse weather conditions, only one of the routes (Tapio) follows the shortest path, while Teemu and Tuire are both instructed to drive west to the coast (Figure \ref{fig:baseroutes}). This allows them to avoid the over 40 km of road work sites located along the shortest route, between points A-C and C-F (Figure \ref{fig:basemap}). Without these roadwork sites, both Teemu and Tuire would also be routed close to the same route as Tapio. Overall, we can see that road work sites are very commonplace in the dataset. 
On the other hand, both traffic and weather weigh relatively little. The highest traffic weights can predictably be found near Oulu, while most of the area further away from the city is weighted well below the average. 
Weather weights experience even less variation, with a difference of only 0.1883 between the maximum and minimum values. A hot spot can be seen some 25 km northeast of Ylivieska, surrounding point F. The weather station mainly responsible for the data in this area has somewhat limited attribute availability. As a result, only 10 out of 15 weather factors could be calculated for this station, putting more emphasis on the 10 factors that were determined. For example, as neither precipitation type, precipitation intensity, nor visible distance could be determined, the visibility factor of the station is calculated based on the dew/frost point and the relative air humidity. 

\begin{figure}
    \centering
    \includegraphics[width=1\linewidth]{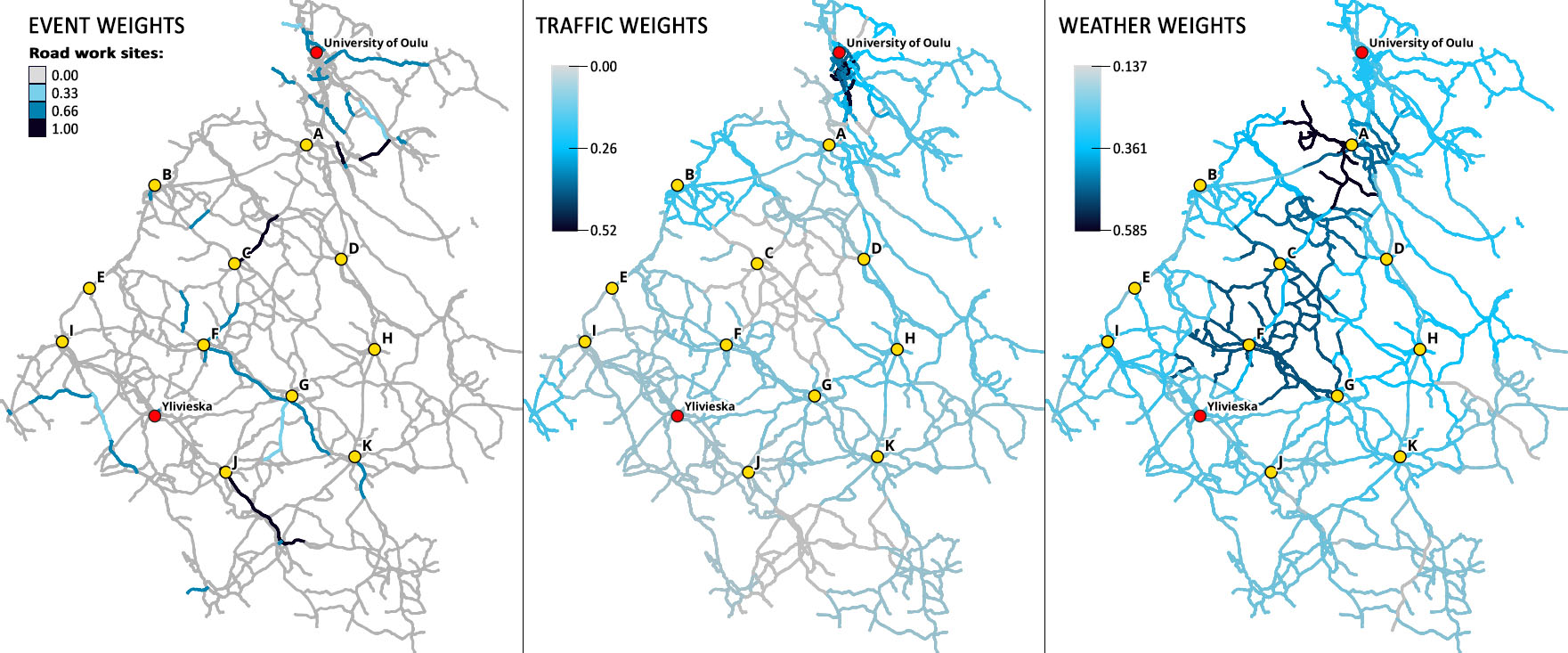}
    \caption{Weight distribution in the thunder dataset for events, traffic measurements, and weather in the area of interest.}
    \label{fig:thundermap}
\end{figure}

The second data set was recorded only three days after the first one, which meant that the majority of road work sites remained the same as in the previous dataset (compare left maps in Figures \ref{fig:basemap} and \ref{fig:thundermap}. 
The weather, on the other hand, was very different. The weight range alone has doubled, as the roads hit by the thunderstorm can be seen in near black in Figure \ref{fig:thundermap} (center). Traffic weights also increased from the previous dataset, particularly around Oulu and along the national road 4. This is likely since the dataset was recorded on a Friday afternoon around 15:45 when many people depart from their place of work. As a result of this, we can see a slight difference in Teemu's route: instead of driving straight south from the university, he is instructed to take a small detour west to avoid the busy road (left in Figure \ref{fig:thunderroutes}). Otherwise, his route will stay the same as previously, avoiding the same 40 km of road work sites between points A-C-F that were present in the first dataset. Tapio, who is less bothered by the change in traffic and completely unbothered by weather, is still routed along the same, shortest route via points University-A-C-F-Ylivieska as previously (center in Figures \ref{fig:baseroutes} and \ref{fig:thunderroutes}).

Tuire's route, on the other hand, experiences the most significant change (right in Figure \ref{fig:thunderroutes}). To avoid both road works and the areas affected by thunder, her route diverts inland, from near point A towards D instead of B as in the previous dataset. 
The first half of her route closely follows the alternative route suggested by Google Maps (Figure \ref{fig:mapsroutes}), but instead of turning west at point D, the route continues south to H and K, from where it turns towards point I and ultimately Ylivieska. This avoids road work sites along the route suggested by Google Maps, in addition to the somewhat heavier weather weights on the shorter route. The weights on the shorter route are derived from the environmental data of weather stations located on the chosen route - this is an example of how the system prefers (perhaps unjustly) roads with complete data. If the full data from the stations along the chosen route were used, the shorter route from point H to G to Ylivieska would be recommended instead, despite the minor road work sites.

\begin{figure}
    \centering
    \includegraphics[width=1\linewidth]{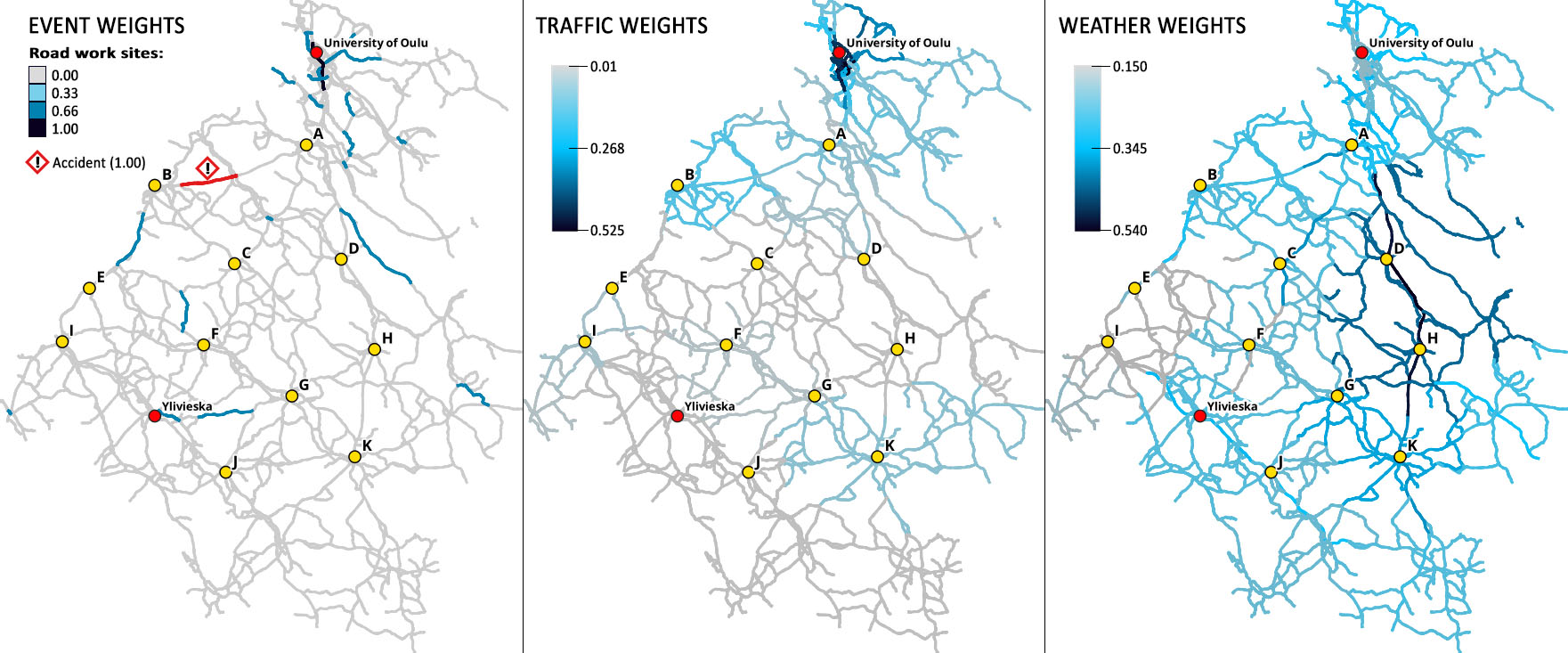}
    \caption{Weight distribution in the October dataset for events, traffic measurements, and weather in the area of interest.}
    \label{fig:octobermap}
\end{figure}

The last set of routes was created for a chilly October morning, with much of the inland part of the area of interest having experienced some light snowfall overnight. With two months since the first dataset, most road work sites differ. This is the only dataset with an accident between points A and B. As in the previous sets, traffic is mainly concentrated around Oulu, while the southern part of the area of interest experiences weights below the average. 
A part of the area, particularly around points D and H, was still experiencing snowfall in the morning, causing poor visibility and surface conditions. This can be seen reflected in the high weather weights in Figure \ref{fig:octobermap}. Therefore, the shortest route from the University via points A-C-F to Ylivieska has been calculated for all users \ref{fig:octoberroutes}. There are some small differences between the users, e.g., Teemu's route avoids traffic initially by taking a small detour inland. The routes created for this dataset most closely resemble the route suggested by Google Maps (Figure \ref{fig:mapsroutes}). 

As no empirical experiment involving humans was conducted, it is hard to assess how satisfied real individuals would be with their routes. Based on the information that can be derived from the data by visually comparing the distribution of weights in the area of interest (Figures \ref{fig:basemap}, \ref{fig:thundermap}, \ref{fig:octobermap}) together with the user descriptions given in Section \ref{sec:eval}, we can make some estimates. Teemu, who wanted to avoid traffic and road work sites, was consistently routed away. When the shortest route was "blocked" by road work sites, he was instead directed to drive along the coast, as was the case with the baseline dataset and the thunder dataset (left in Figures \ref{fig:baseroutes} and \ref{fig:thunderroutes} respectively). Similarly, his route avoids roads with high traffic weights, however as these high weights are mostly present near the city of Oulu, their effect on the overall route is smaller. Based on this, Teemu could be somewhat satisfied with his routes.

Tuire also wanted to avoid road work sites and adverse weather conditions. As a result, in the baseline dataset, her route was diverted to the coast to avoid the road work sites on the shortest path, similar to Teemu's (Figure \ref{fig:baseroutes}, left and right). In the thunder dataset, her route detoured inland to avoid the road work sites and the thunderstorm closer to the coast. As a result of this detour, her route is significantly longer than the shortest route but follows the preferences assigned to her. Therefore, it would be reasonable to assume that Tuire would be mostly satisfied with her routes.

Tapio, on the other hand, only wanted the shortest path. Visually, the paths generated for him (center in Figures \ref{fig:baseroutes}, \ref{fig:thunderroutes}, and \ref{fig:octoberroutes}) are close to the shortest, but compared to the route suggested by Google Maps, the route differs. Rather than following the shortest path, the generated route inexplicably deviates from it by including multiple unnecessary detours onto adjacent roads, only to return to the original road. Most of these detours are relatively short, ranging from 20 m to some hundreds of meters of added distance, but the longest one, right before Ylivieska, is over 9.3 km long, adding almost 5 km to the overall length of the path. Overall, the route generated for Tapio is nearly 10 km longer than the route suggested by Google Maps at 144.6 km, compared to 135 km in Google Maps. Therefore, it could be assumed that Tapio would be somewhat dissatisfied with his route.
Curiously enough, the route generated for Tuire in the last dataset (yellow on the right in Figure \ref{fig:octoberroutes}) follows more closely along the path by Google Maps, lacking most of the detours present in Tapio's route. As a result, the length of this route comes to 135.7 km. 

This leads us to believe that the "issue" might lie with the lack of normalizing the weights of road segment lengths, which was done with all other data. The lengths can vary significantly, and therefore, the weights can vary. Additionally, the dataset used to create the road network did not distinguish between roads meant for motor vehicles and pedestrians, sometimes leading to routes that would not be permissible in the real world due to traffic laws.
\section{Discussion}

The nine routes generated with the system generally met our expectations. The routes differed due to different preferences when they were expected to be (e.g., in the thunder dataset) but similar when there was little reason to diverge (e.g., in the early winter dataset). Based on the results, it can be concluded that the system works well for avoiding poor weather, congestion, and road work sites. Still, surprisingly, it struggles with finding the geographically shortest route when the other factors are not given much weight. It is not known what exactly causes this, although it is likely to be related to the fact that, unlike the other factors, the lengths of road segments were not normalized. 

It is also possible that as Dijkstra's algorithm only takes one single value as weight, and we had to reduce the dimensions of the attributes involved, it could impact the accuracy of the route prediction. We are planning to try GNN as a replacement for this Dijkstra's based system, but it is possible that it requires too much computing power and time for it to be applicable in an automotive setting. The benefits of it could be that it could take advantage of the graph nature of the dataset, be more adaptable, and learn better about the driver, thus providing better route recommendations. 

Additionally, the system has some limitations, as it is meant to provide a proof-of-concept of how the data could be utilized.
The system lacks consideration for the time dimension: conditions on the road change over time, especially as the traveled distance grows. In real life, this could mean that by the time the user has traveled further along the route, the conditions based on which the route was chosen have altered, and another route would, in fact, be more optimal. This issue could be mitigated in one of two ways: recalculating the route as the trip progresses or predicting the conditions farther away from the point of origin. The API readily offers weather forecasts that could have been used for this purpose; however, utilizing these would require the calculation of travel times, which was deemed outside this project's scope. Predicting traffic volumes and speeds would also require travel time estimates, as well as the use of historical data to discover patterns in congestion. Similarly, this was left for future work. 
The lack of time dimension also meant that we did not consider the effects of the dark time of the day. It is known that traveling at night or otherwise in the dark introduces risks that are not equally pronounced in a daylight setting. By considering the time dimension, we could prioritize shorter routes, particularly in situations where the sun is about to set during the journey, or with additional information from the Finnish Transport Infrastructure Agency, when driving at or after twilight, for example, roads with street lights could be prioritized for the safety of the user. This could also be the source of another preference weight, as some people dislike driving in the dark significantly more than others.

The system, in its current state, does not operate in real-time. The MQTT listener must first collect the data from measurement stations, which can take up to 5 minutes due to the update interval. Additionally, maintenance alerts must be queried and filtered to determine those currently affecting traffic. While all of this could be streamlined and "automated" to run at the click of one button, these improvements were left for future work. 
"Real life" produces many other hindrances as well.
As seen in Figure \ref{fig:areaofinterest}, measurement stations can be sparse. Therefore, no two data sets can be considered equally reliable. For example, readings from a station close by on the same road segment as the inspected segment are more reliable than those collected from a station 16 km away. This is especially true for traffic measurements, as well as road surface conditions, which are also affected by passing traffic and maintenance operations. This effect is less prominent with environmental weather factors, such as temperature and winds. This is why, in our system, we differentiated between the weather data used for road segments with their assigned weather station on the same road and those on different roads. The downside of this approach is that the system tends to assign higher weights to the roads that do not have a weather station on them. This, of course, is very situation-specific and sometimes goes the other way around. Regardless, this causes an unjust inequality between the roads. In future work, we shall utilize more advanced strategies, such as interpolating the weather data based on values from multiple stations. With traffic data, using the data from adjacent roads to assess the congestion on others is a very simplistic approach and hardly the most reliable. Roads of different sizes have very different capacities, not to mention traffic patterns, making these generalizations akin to guesswork. For more accurate estimates of conditions on other roads, in future work, we would like to use historical data from the Finnish Transport Infrastructure Agency to get daily averages and observed trends in traffic patterns to aid with these estimates.

Interpolating from nearby stations and historical data from a similar situation could also be used when certain data is missing from a weather station or if a station is missing completely. The current practice of manually selecting a secondary station only when there is no data from a certain station lacks robustness. The system's reliability could be greatly improved by acquiring data to fall back on should the station not produce any specific data. This would require first collecting historical data. To our knowledge, historical data from weather stations or roads in our area of interest is not publicly available. These considerations are important because, at times, there are great differences in attributes present at different stations, making comparisons between them and the road segments assigned to them unjust. Additionally, our system treats all stations equally without considering their differences. For example, the maximum visible distance was set to 10 km despite a few stations only perceiving distances up to 2 km, even in pristine conditions. 

To aid with the geographical sparsity of the measurement stations, we intend to utilize many road cameras in the future to fill in some of the gaps. Of course, the data that can be derived from images vastly differs from that of detailed weather or traffic measurements. However, cameras can still be valuable in recognizing the main characteristics of conditions, as was discussed in related works. In an ideal situation, many more measurement stations would be installed, particularly on roads that currently do not have any. However, measurement stations are certainly expensive to both install and maintain. Therefore, it is not likely that their numbers will increase significantly any time soon. As such, it is better to focus on the hardware currently in place.

Even with all available measurement data, the approach of using individual, isolated attributes to determine the weather weights is perhaps too simplistic. In reality, many factors influence one another. For example, rain is rarely associated with extremely adverse weather, particularly if the temperature is above freezing unless the precipitation intensity is significant. On the other hand, even light rain on an icy road surface can make the conditions on the road near disastrous. 
Additionally, the many attributes lack sufficient documentation, meaning that some educated guesses had to be made to determine the ranges of values. Some attributes could also not be used as there was no reliable way to determine the meaning behind the numerical values. Deciphering these attributes would be beneficial as they could provide data to fall back on if some of the attributes used in the work were not available.
In future work, we will incorporate such factors that influence conditions in a way that differs from their individual effect. This will allow us to create more appropriate routes, particularly for individuals with strong preferences to avoid adverse weather conditions.

On the topic of weather conditions, in the future, we hope to provide more detailed preference vectors that enable us to accommodate more specific preferences regarding weather. For example, this would allow us to provide users who prefer salting agents on roads with more optimal routes. In future work, we shall also introduce preference vectors derived from the true preferences of real-world individuals, as in the current work, the preference vectors were created based on fictional individuals. We also hope to receive feedback from real people about the optimality of the routes generated, which would greatly improve our understanding of the performance of the weight definition and the overall system.

\section{Conclusions}
Route creation services often give the shortest route as a route recommendation, while the most optimal route for a specific driver can differ vastly from this. There is, thus, a need for more personalized route recommendation engines. These recommendation engines require data to create the optimal suggestions. The Finnish Digitraffic provides an open road data interface for accessing data from a nationwide road sensor network of 1,814 sensor stations, which has over a thousand different attributes. We utilized the data available from Digitraffic to create more personalized route recommendations. We presented an extensive analysis of the data available from Digitraffic and a description of how it was preprocessed, fused, and cleaned for our use case. 

Out of the data available from Digitraffic, we utilized the road weather, traffic, and road event data for the route recommendation system. This data, together with the preference vectors we created, were used to generate the weight vector, which was inserted into Dijkstra's algorithm. In this initial iteration of the recommendation engine, Dijkstra’s algorithm served as a baseline route generation method, allowing us to concentrate on integrating the Digitraffic data and preference vectors. The system was then evaluated with three different driver profiles, which we created in three situations with varying weather conditions and other road events.

Our route recommendation engine seemed to generate routes that matched well with the profiles created for our example drivers. In our future work, we plan to replace Dijkstra's algorithm with a GNN to further improve the personalization capabilities and to estimate its suitability for our use case. Using a GNN would require data on real driver preferences, which we plan to collect. Potentially, other machine learning algorithms could also be tested to evaluate the best algorithm for learning driver preferences and providing the most optimal personalized route recommendations. 

\section*{Acknowledgments}
The work has been supported by the EU HORIZON projects CHIPS-JU CIA FEDERATE (grant number 101139749) and CHIPS-JU RIA HAL4SDV (grant number 101139789), Business Finland projects 6G Visible (grant number 10743/31/2022) and HAL4SDV national funding (grant number 7655/31/2023), and the Finnish Research Council project Northern Utility Vehicle Laboratory Consortium GO!-RI (grant number 352726).

\bibliographystyle{plain}
\bibliography{references}

\end{document}